%% file: main.tex
\documentclass[acmsmall]{acmart}
\usepackage{graphicx}

\usepackage{amsmath,amssymb,amsfonts}
\usepackage{algorithmic}
\usepackage{graphicx}
\usepackage{todonotes}
\usepackage{caption} 
\usepackage{multirow}
\usepackage{float}
\usepackage{subfig}

\usepackage{amsmath}
\usepackage{dblfloatfix}
\usepackage[redeflists]{IEEEtrantools}
\usepackage{fancyvrb}
\fvset{%
numbers=left
}

\newcommand{\NA}{NA}
\newcommand{\NR}{NR}

\usepackage{pifont}
\newcommand{\cmark}{\ding{51}}%
\newcommand{\xmark}{\ding{55}}%

\AtBeginDocument{%
  \providecommand\BibTeX{{%
    \normalfont B\kern-0.5em{\scshape i\kern-0.25em b}\kern-0.8em\TeX}}}

\usepackage{array}
\newcolumntype{P}[1]{>{\centering\arraybackslash}p{#1}}

\acmJournal{TACO}

\begin{document}

\title{CLARINET: A RISC-V Based Framework for Posit Arithmetic Empiricism}


\author{Niraj N.~Sharma}
\orcid{0000-0003-3311-3442}
\affiliation{
  \institution{Indian Institute of Technology, Bombay}
  \streetaddress{4th Floor, Department of Electrical Engineering}
  \city{Mumbai}
  \country{India}}
\email{nirajns@iitb.ac.in}

\author{Riya Jain}
\email{riyajain78@gmail.com}
\affiliation{%
  \institution{Indian Institute of Technology, Bombay}
  \city{Mumbai}
  \country{India}
}

\author{Madhumita Mohan}
\email{pmmadhumita9@gmail.com}
\affiliation{%
  \institution{Indian Institute of Technology, Bombay}
  \city{Mumbai}
  \country{India}
}

\author{Sachin Patkar}
\affiliation{%
  \institution{Indian Institute of Technology, Bombay}
  \streetaddress{4th Floor, Department of Electrical Engineering }
  \city{Mumbai}
  \country{India}}

\author{Rainer Leupers}
\affiliation{%
 \institution{RWTH Aachen University}
 \streetaddress{Kopernikusstraße 16}
 \city{Aachen}
 \state{NRW}
 \country{Germany}}

\author{Nikhil Rishiyur}
\orcid{1234-5678-9012}
\affiliation{\institution{Bluespec Inc.}
\city{Framingham}
\state{MA}
\country{USA}}

\author{Farhad Merchant}
\affiliation{
  \institution{RWTH Aachen University}
  \city{Aachen}
  \country{Germany}
}

\renewcommand{\shortauthors}{Sharma et al.}

\begin{abstract}
\input{abstract}
\end{abstract}


\maketitle





\section{Introduction}
\input{introduction}

\section{Background and Related Work}\label{sec:rw}
\input{background}

\section{Clarinet}\label{sec:clarinet}
\input{clarinet}

\section{Melodica}\label{sec:melodica}
\input{melodica}

\section{Case Studies}\label{sec:exp}
\input{case}

\section{Conclusion} \label{sec:con}
\input{conclusion}

\bibliographystyle{ACM-Reference-Format}
\bibliography{main}


\end{document}

%% file: abstract.tex
Many engineering and scientific applications require high precision arithmetic. IEEE~754-2008 compliant (floating-point) arithmetic is the de facto standard for performing these computations. Recently, \emph{posit arithmetic} has been proposed as a drop-in replacement for floating-point arithmetic. The posit\texttrademark  data representation and arithmetic claim several absolute advantages over the floating-point format and arithmetic, including higher dynamic range, better accuracy, and superior performance-area trade-offs. However, there does not exist any accessible, holistic framework that facilitates the validation of these claims of posit arithmetic, especially when the claims involve long accumulations (\emph{quire}). 

In this paper, we present a consolidated general-purpose processor-based framework to support posit arithmetic empiricism. The end-users of the framework have the liberty to seamlessly experiment with their applications using posit and floating-point arithmetic since the framework is designed for the two number systems to coexist. Melodica is a posit arithmetic core that implements parametric fused operations that uniquely involve the quire data type. Clarinet is a Melodica-enabled processor based on the RISC-V ISA. To the best of our knowledge, this is the first-ever integration of quire with a RISC-V core. To show the effectiveness of the Clarinet platform, we perform an extensive application study and benchmark some of the common linear algebra and computer vision kernels.
We emulate Clarinet on a Xilinx FPGA and present utilization and timing data. Clarinet and Melodica remain actively under development and is available in open-source for posit arithmetic empiricism.

%% file: introduction.tex
High precision arithmetic is necessary for several scientific and engineering applications~\cite{higham1}. These applications require high precision arithmetic hardware for accurate and efficient execution. Floating-point arithmetic hardware has been challenging to design due to the intricacies involved in staying within the limits of the desired area, and power budget~\cite{leeser01}\cite{florent01}. Specifically, reproducing results across floating-point unit (FPU) implementations has been a challenge due to the requirements posed by the IEEE~754-2008 (floating-point) standard. Recently, researchers and computer architects have either compromised on compliance to the standard or devised alternate formats to overcome the design challenges~\cite{Gustafson2017}~\cite{bf16}. \textit{Posit} is one such data representation proposed by John L.~Gustafson in the year 2017, which aims to overcome shortcomings of the floating-point format~\cite{Gustafson2017}. 

The posit arithmetic and data representation claim simpler hardware to implement, higher dynamic range and numerical accuracy with comparable area and energy footprints as advantages over the floating-point arithmetic and format. In general, $n$-bit posit has a higher dynamic range compared to $n$-bit floating-point, and $n$-bit floating-point arithmetic can be replaced by $m$-bit posit arithmetic units where $m<n$~\cite{Gustafson2018}. The posit representation is a super-set of the floating-point format and can serve as a drop-in replacement. It is also shown that the posit format is robust and reliable compared to its floating-point counterpart for single and double bit-flips~\cite{positreliability}. 

Due to the advantages of the posit number system, several academic and industrial research labs have started exploring and studying applications that can benefit from posits. The \emph{SoftPosit} library emulates posit arithmetic and supports early-stage numerical investigation of applications in software~\cite{softposit}. However, no such framework exists to evaluate posits from a latency, resources and operating frequency perspectives on actual hardware. There is a need for an easily reconfigurable hardware platform for early-stage design space exploration of posit arithmetic for various applications. With its ever-increasing popularity and a conducive open-source ecosystem, we believe that RISC-V~\cite{rvisa} is an excellent vehicle to have such a framework supporting posit arithmetic empiricism. We chose the BSV high-level HDL~\cite{bsv-git} as the implementation language to enable rapid design space exploration through an easy reconfiguration of the hardware platform. The major contributions of this paper are:

\begin{itemize}
    \item We present \emph{Clarinet}; a floating-point and posit arithmetic enabled CPU-based framework for numeric empiricism. Clarinet is based on the RISC-V ISA (with custom instructions for posit arithmetic), and is derived from the open-source \emph{Flute} core developed by \emph{Bluespec Inc}~\cite{flute}. The Clarinet framework also features a customized RISC-V \texttt{gcc} tool-chain to support the new instructions.
    \item We present \emph{Melodica}, a parameterizable posit arithmetic core which supports fused operations that involve the quire data type, and type-converters between floating-point, posit and quire data representations. 
    \item Through Clarinet, we also present a new usage model where posits and floating-point can coexist as independent types cleanly, allowing applications to be ported more easily to posits when they offer an advantage.
    \item Finally, we investigate applications in the domain of linear algebra and computer vision to show the effectiveness of Clarinet as an experimental platform.
\end{itemize}



We support fused multiply accumulate (and subtract), and fused divide accumulate (and subtract) with quire to carry-out experimental studies for our applications. To the best of our knowledge, this is the \emph{first-ever} quire enabled RISC-V CPU. The organization of the paper is as follows. In Section~\ref{sec:rw}, we discuss posit, quire and float formats, the Flute core, and some of the recent implementations of posit arithmetic. Clarinet is described in Section~\ref{sec:clarinet}, and Melodica in Section~\ref{sec:melodica}. Application analyses and benchmarks are presented in Section~\ref{sec:exp}. Experimental setup and results are discussed in Section~\ref{sec:res}. We conclude our work in Section~\ref{sec:con}.

%% file: background.tex
\subsection{Background}
\subsubsection{Posits} \label{sec:background:posits}
A posit number is defined by two parameters: the width of the posit
number, \texttt{N}, and the maximum width of the exponent field, \texttt{es}. One of
the important advantages of the posit number format is that we can vary
\texttt{es} to trade-off between greater dynamic range (larger \texttt{es}) and greater
precision (smaller \texttt{es}). The posit format has four fields: 

\begin{figure}[!t]
\centering
 \includegraphics[width=\columnwidth]{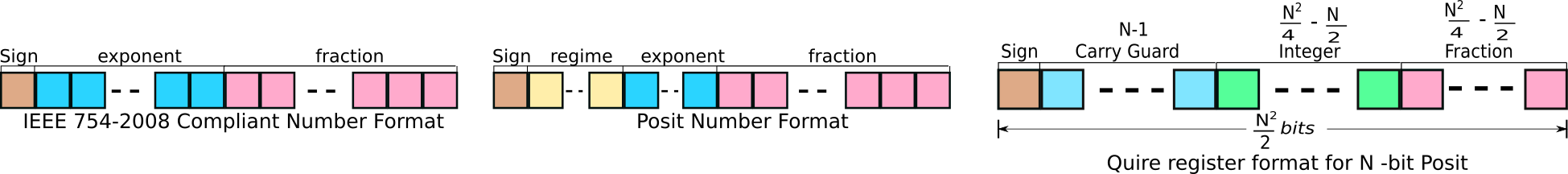}
\caption{IEEE~754-2008 floating-point, posit, and quire register format for \texttt{N}-bit posits}
\vspace{-5mm}
\label{fig:quireexample}
\end{figure}

\begin{itemize}
    \item Sign (\texttt{s}): The MSB of the number. If the bit is set, the posit value
       is negative. In this case all remaining fields are represented
       in two's complement notation.
    \item Regime Field (\texttt{r}): The regime is used to compute the scale factor, \texttt{k}. In a posit number this field starts just after the sign bit
       and is terminated by a bit opposite to its leading bits. The
       computation of \texttt{k} is as per the equation~\ref{eqn:k}, where \texttt{r} is the
       number of leading bits in the regime.
    \item Exponent Field (\texttt{exp}): The exponent begins after the regime field and the maximum width of the exponent field is \texttt{es}.
    \item Fraction Field (\texttt{f}): The remaining number of bits after the
       exponent make up the fraction. The fractional field
       is preceded by an implied hidden bit which is always 1.
\end{itemize}

\noindent For a number represented in the posit format, its value is as per the equation~\ref{eqn:posit-val}.

\begin{equation}
   k = \begin{cases}
   r - 1, & \text{if regime has leading ones}\\
   -r,    & \text{if regime has leading zeros}
    \end{cases}
    \label{eqn:k}
\end{equation}

\begin{equation}
value = (-1)^{s}*(2^{2^{es}})^{k}*2^{exp}*1.f
\label{eqn:posit-val}
\end{equation}



Posits do not have a representation for NaNs, or separate representations for $\pm 0$ and $\pm \infty$. Posits recognize only two special cases -- zero and not-a-real (NaR), and support one rounding mode Round-to-Nearest-Even (\texttt{RNE}). Posit number system shows better accuracy around 1 than floating-point of the same size~\cite{florent02}. Using 8-bit posits as an example. \texttt{0000\_0000} represents \texttt{0}, \texttt{1000\_0000} represents \texttt{$\pm \infty$}, and the rest of the combinations can be derived from the equation~\ref{eqn:posit-val}. Posit and floating-point formats are depicted in Fig.~\ref{fig:quireexample}.

The \emph{quire} is a fixed-point register that serves the purpose of accumulation like a Kulisch accumulator~\cite{kulisch1}. The quire for a given posit-width is sized to represent the smallest posit squared, and the largest posit squared without any overflow. When the quire is used as an accumulator for a series of steps, it allows computation without intermediate rounding. The size of an \texttt{N}-bit quire is determined by $\frac{N^2}{2}$ where \texttt{N} is the posit number width (Fig.~\ref{fig:quireexample}). 

\subsubsection{Flute - A RISC-V CPU}
\emph{The Flute} is an in-order open-source CPU based on the RISC-V ISA, implemented using the BSV HL-HDL. The Flute pipeline is nominally 5-stages but longer for instructions like memory load-stores, integer-multiply, or floating-point operations. The core supports high-level parameters that allow it to be configured to operate at 32-bit (RV32) or 64-bit (RV64) and support different variants of the RISC-V ISA~\cite{rvisa}. The Flute core also supports a memory management unit (MMU) and is capable of booting the Linux operating system. The pipeline stages in Flute are:

\begin{description}
\item[F:] Issue fetch requests to the instruction memory. The fetch stage can also handle compressed instructions.
\item[D:] Decode the fetched instruction. Checks for illegal instructions.
\item[E1:] The first execution stage. Reads the register files or accept forwarded values from earlier instructions. Execute all single-cycle opcodes meant for the integer ALU. Branches are resolved here. Discard speculative instructions.
\item[E2:] Execute multi-cycle operations, including floating-point operations. Multi-cycle operations are dispatched to their individual pipelines from this stage. If the instruction was executed in E1, this stage is just a pass-through.
\item[WB:] Collects responses from various multi-cycle pipelines, handle exceptions and asynchronous events like interrupts, and commit the instruction.
\end{description}

\subsection{Related work}

\begin{table}[!b]
\tiny
\centering
\caption{Posit Arithmetic Implementations in the Literature}
\begin{tabular}{|p{1.7cm}| P{0.8cm} | P{1.0cm}| P{1.1cm}| P{1.1cm}| P{1.7cm}| P{0.8cm}| P{0.8cm}| P{0.8cm}|}
\hline
Impl. & Fully parametric & Application\newline specific  & Application \newline study & RISC-V \newline integration  & Posit custom\newline instruction support & Open source & SoftPosit\newline porting & Quire\newline support \\
\hline
Uguen~\cite{florent03}       & \cmark & \xmark & \xmark & \xmark & \NA & \cmark & \NA & \cmark  \\
Jaiswal~\cite{manish3}       & \xmark & \xmark & \xmark & \xmark & \NA & \cmark & \NA & \xmark  \\
PAU~\cite{Gustafson2018}     & \cmark & \xmark & \xmark & \cmark & \NA & \xmark & \NA & \xmark  \\
Lu~\cite{posittraining1}     & \xmark & \cmark & \cmark & \xmark & \NA & \cmark & \NA & \xmark  \\
Adaptive Posit~\cite{adaptiveposit}  & \cmark & \cmark & \cmark & \xmark & \NA & \xmark & \NA & \xmark  \\
SmallPositHDL~\cite{smallposithdl}  & \xmark & \cmark & \xmark & \xmark & \NA & \cmark & \NA & \cmark  \\
Deep PeNSieve~\cite{deeppensieve}  & \cmark & \cmark & \cmark & \xmark & \NA & \cmark & \NA & \cmark  \\
Jaiswal~\cite{manish2}       & \cmark & \xmark & \xmark & \xmark & \NA & \cmark & \NA & \xmark  \\
Cheetah~\cite{langroudi2019cheetah} & \xmark & \cmark & \cmark & \xmark & \NA & \xmark & \NA & \cmark  \\
Deep Positron~\cite{deeppositron} & \xmark & \cmark & \cmark & \xmark & \NA & \cmark & \NA & \cmark  \\
ExPAN(N)D~\cite{expand1}     & \xmark & \cmark & \cmark & \xmark & \xmark & \cmark & \xmark & \xmark  \\
{\bf Melodica}                     & \cmark & \xmark & \cmark & \cmark & \cmark & \cmark & \cmark & \cmark \\ \hline
\multicolumn{9}{|c|}{RISC-V Integration} \\ \hline
PERI~\cite{Tiwari2019}       & \cmark & \xmark & \xmark & \cmark & \xmark & \xmark & \xmark & \xmark  \\
PERC~\cite{perc}             & \cmark & \xmark & \xmark & \cmark & \xmark & \cmark & \xmark & \xmark  \\
CRISP\footnote{Based on publicly available data}~\cite{crisp1}               & \xmark & \xmark & \xmark & \cmark & \xmark & \xmark & \xmark & \xmark  \\
Saxena~\cite{rvposit1}       & \cmark & \xmark & \cmark & \cmark & \xmark & \xmark & \xmark & \xmark  \\
POSAR~\cite{posar}       & \cmark & \xmark & \cmark & \cmark & \xmark & \cmark & \xmark & \xmark  \\
{\bf Clarinet}                     & \cmark & \xmark & \cmark & \cmark & \cmark & \cmark & \cmark & \cmark  \\
\hline
\end{tabular}
 \label{tab:posit_literature}
\end{table}

Since the inception of posit data representation and arithmetic, there have been several implementations of posit arithmetic in the literature. 
In~\cite{manish1}, the authors have covered the design of a parametric posit adder/subtractor, while in~\cite{manish2}, the authors have presented parametric designs of float-to-posit and posit-to-float converters, and multiplier along with the design of adder/subtractor. The \emph{PACoGen} open-source framework in~\cite{manish3} generates pipelined adder/subtractor, multiplier, and divider. PACoGen is capable of generating the hardware units that adapts precision at run-time. A more reliable implementation of a parametric posit adder and multiplier generator is presented in~\cite{Gustafson2018}. However, the implementation in~\cite{Gustafson2018} is not pipelined resulting in low operating frequency for large bit-widths. \emph{Cheetah} presented in~\cite{langroudi2019cheetah} discusses the training of deep neural network (DNN) using posits. We believe that the architecture presented in~\cite{langroudi2019cheetah} is promising and some of the features can be incorporated in Melodica in the future. Apart from the mentioned efforts, there have been several other implementations of
posit hardware units~\cite{iscas1}\cite{zhang1}. 

More recently, there has been an interest to integrate posit numeric units with RISC-V processors and demonstrate applications with posit arithmetic. \emph{PERI}~\cite{Tiwari2019} integrates a posit numeric unit as a functional unit with the Shakti C-Class RISC-V processor, and \emph{POSAR}~\cite{posar} integrates a parameterized posit arithmetic unit with a Rocket Chip-based RISC-V core. These implementations do not support quire and redirect
floating-point instructions and operands to the new posit numeric unit. Coupled
with the hardware changes, a gcc back-end function which represents real number
variables in posit notation, makes computations using posits transparent to the
programmer. PERI, POSAR, PERC, CRISP, and Saxena~\cite{rvposit1} cannot support the breadth of experimentation possible
in Clarinet because (a) only 32-bit and 64-bit posits may be used, (b) floats and
posits may not co-exist in the same program, and (c) quire computing is not
supported. 

None of the previous efforts have resulted in a complete computing environment
that allows hardware-software posit arithmetic empiricism. Further, they do not
include an easy-to-use software framework that allows floating-point and posit
types to cohabit in an application. Clarinet has been created to serve
as an open-source test-bed for hardware-software experimentation around posits. By
allowing floating-point and posits to coexist in an application,
Clarinet uniquely allows researchers to make trade-offs between hardware costs,
latency and precision. Table~\ref{tab:posit_literature} summarizes the posit
arithmetic implementations and features such as parameterization,
application specificity, RISC-V integration, quire support, and SoftPosit
porting. The only feature that is not supported in Clarinet is application
specificity. Since the Melodica unit is fully parametric, it is practicable to
generate the application-specific instances of the unit for domain-specific
computing platforms which is true for the other literature works that are
parametric.


%% file: clarinet.tex
The system comprises two main components -- \emph{Melodica}, a
\emph{parameterizable} posit arithmetic unit that supports quire-based
arithmetic described in Section~\ref{sec:melodica}, and
\emph{Clarinet}, a RISC-V CPU that is enhanced with special
instructions for posit arithmetic and a dedicated Posit Register
File (PRF).

\subsection{Clarinet organization -- a type-centric approach}
In addition to the existing floating-point types (float and
double) which exist in the RISC-V architecture, Clarinet
introduces two new types for computation with real numbers --
posit and quire. The posit type is parameterized by two numeric values --
\texttt{N} and \texttt{es} as described in
Section~\ref{sec:background:posits}. The Posit Register File (PRF) holds
values of posit type. Values of posit type may be read from and written to
memory, and may be accessed directly by the programmer. The quire register is
an accumulator inside Melodica, which accumulates values of the quire type.
The quire type is paramterized by \texttt{N} as described in
Section~\ref{sec:background:posits}. The quire values reside only inside the
quire register and may not be accessed directly by software. A quire value
however may be converted to a posit value and made available to the programmer
via the PRF.

\begin{figure}[!t] 
\centering
\subfloat[Clarinet pipeline]{
\includegraphics[width=0.44\columnwidth]{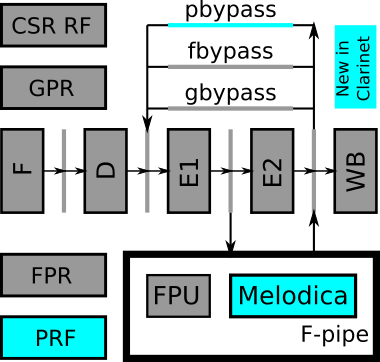}
\label{fig:clarinet-bd}}
\subfloat[New Clarinet instructions]{
\includegraphics[width=0.35\columnwidth]{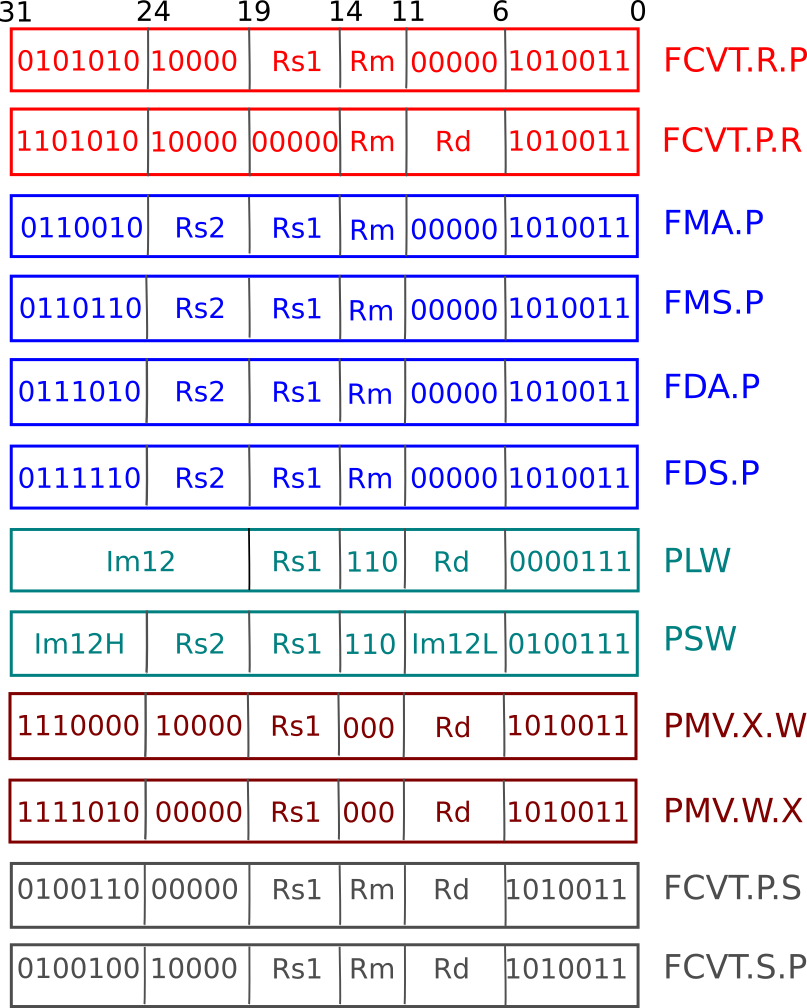}
\label{fig:new-instr}}
\caption{Clarinet}
\vspace{-5mm}
\end{figure}

Clarinet's organization is illustrated in Fig.~\ref{fig:clarinet-bd}. 
The starting point for Clarinet was a Flute CPU core, configured with the RV32IMAFC variant of the RISC-V ISA~\cite{rvisa}. 
Clarinet integrates Melodica as a functional unit parallel to the existing floating point unit.
The module \texttt{mkFBox\_Core}, integrates the existing floating point core, and the new Melodica core. 
Decode logic directs the new instructions described in Section~\ref{sec:custom-instr} to Melodica, while all other floating point instructions continue to be serviced by the FPU, responses from Melodica are routed back to the Clarinet pipeline.

\subsection{New instructions} \label{sec:custom-instr}
\subsubsection{Quire-based computing} \label{sec:quire-computing}
Computation using the quire accumulator in Melodica can be visualized as
comprising of the following three operations: (i) initialize the quire
register (accumulator), (ii) compute into the quire, and (iii) read the result
from the quire by converting into a posit value. Six new instructions have
been been added to support these operations. The instruction \texttt{FCVT.R.P}
initializes the quire with a posit value from the PRF. The instructions
\texttt{FMA.P}, \texttt{FMS.P}, \texttt{FDA.P}, and \texttt{FDS.P} perform
fused multiply (M) or divide (D) addition (A) and subtraction (S) computations
using two posit values from the PRF. The result accumulates into the quire
register. The instruction \texttt{FCVT.P.R} converts the value in the quire
register into a posit value destined for the PRF.

\subsubsection{Interfacing with memory}
Two instructions -- \texttt{PLW} and \texttt{PSW} provide memory access to
load and store posits. These instructions use the GPR to compute the memory
address, and the PRF for the data. In addition, two new move instructions,
\texttt{PMV.X.W} and \texttt{PMV.W.X} move data between the GPR and PRF.
Unlike conversion instructions, the move instructions do not interpret the
data present in the source register. The PMV instructions were introduced to
allow referencing of posit data structures using integer types well-supported
by the C compilers like \texttt{int}, \texttt{short} and \texttt{char}. As the
move instructions do not reinterpret the source data, these instructions have
low overhead in terms of computation and implementation costs.

\subsubsection{Type converters}
One of the design considerations for Clarinet was to create a CPU which
allowed computing using both floating-point and posit values. For legacy
applications, programmers may want to enable posit arithmetic for loops or
kernels to benefit from the advantages of posits, while retaining floating
point arithmetic for the overall application. Two type-converter instructions,
\texttt{FCVT.P.S} and \texttt{FCVT.S.P}, convert between posit and
floating-point values, using the PRF and FPR respectively. These instructions
can be visualized to be book-ends for a posit computation kernel as described
in~\ref{sec:quire-computing}.

\subsubsection{Instruction encoding}
Bit representations of the new instructions are in
Fig.~\ref{fig:new-instr}. All the instructions except the load-store
instructions belong to the \texttt{R}-format type of the RISC-V ISA. The
\texttt{PLW} instruction is in the \texttt{I}-format, while the \texttt{PSW}
instruction is in the \textbf{S}-format. The instructions reuse existing major
opcode values as defined in ~\cite{rvisa} (bits 6:0). In order to handle the
new posit types, a binary encoding \texttt{10} was introduced for the two-bit
\texttt{fmt} field. In \texttt{R}-format instructions, these bits occupy bits
26:25 of the instruction. In the four compute instructions, this field
indicates the type of the source operands.  The type converter instructions
and quire initialize/read instructions require both source and result
encodings. The \texttt{fmt} serves as the encoding for the result's type,
while the five-bit \texttt{rs2} field serves as the encoding for the source's
type. New \texttt{rs2} encodings have been introduced for posit values
(\texttt{0x10}) and for quire values (\texttt{0x11}). Due to a lack of free
encoding space in the \texttt{fmt} field, a two-bit code was not assigned for
quire values. Differentiation between \texttt{fcvt.r.p} and \texttt{fcvt.p.r}
instructions is done by looking at a combination of \texttt{fmt} and
\texttt{rs2} fields. If the source operand is a quire value (\texttt{rs2}
equals \texttt{0x11}), it is interpreted as a \texttt{fcvt.p.r} instruction,
and if the source and destination operands are posit values (\texttt{rs2}
equals \texttt{0x10} and \texttt{fmt} equals \texttt{10}), it is interpreted
as a \texttt{fcvt.r.p} instruction. For the load and store instructions, the
\texttt{Rm} fields differentiates a posit load or store from the
floating-point loads and stores. The \texttt{Rm} encoding of \texttt{110}
indicates that the load or store uses the PRF. For the move instructions, the
source register file is encoded in the \texttt{rs2} field while the
destination register file is encoded in the \texttt{fmt} field.

The decision to add new instructions instead of reusing existing opcodes
belonging to the \texttt{F} subset of the RISC-V ISA was driven by two
requirements -- integrating quire functionality (whose equivalent does not
exist in the RISC-V floating-point ISA), and type-converter instructions that
would allow posit and floating point values to coexist in an application as
independent types.

\subsubsection{Coexisting posits with floating-point}
The type-converter instructions allow existing programs which use
floating-point arithmetic to benefit from the use of posits and quire (greater
dynamic range or accuracy) by converting certain computation kernels to use
the quire, while retaining the rest of the computation in floating-point. From
our experiments as illustrated in ~\ref{sec:soft-error-analysis}, we observe
that applications see significant reductions in normalized error through the
introduction of quire-based accumulation even when most of the computation
remains in floating-point. When an application can benefit from the use of
posits (be it greater dynamic range or accuracy), the type-converter
instructions allow the user to convert a part of the computation to posits and
accumulate into the quire register. In order to do so, they would first need
to convert their intermediate floating-point data to posits using the
type-converter instructions, before executing one of the compute instruction
that accumulates into the quire. Eventually, the results are converted back to
the floating-point format for further processing.

Using type-converter instructions to insert posit and quire-based computation
into an existing application, introduces an overhead in terms of the number of
instructions that need to be executed to complete the overall computation as
illustrated in the code example on section~\ref{sec:clarinet-eg}. Since this
overhead is present in each iteration of the inner compute loop, for a loop
with \texttt{N} iterations, there are $2N$ extra instructions for type
conversion from floating-point to posit. However, for new application which
are built using posits, there is no overhead in terms of the number of
instructions executed as illustrated in section~\ref{sec:clarinet-eg}.

\subsubsection{Modifying the RISC-V gcc compiler} \label{sec:compiler}
In order to compile programs that uses the new instructions, and work with the
newly introduced PRF, we made modifications to the RISC-V gcc assembler. This
allowed us to compose assembly programs that uses the posit new instructions.
However, we did not extend these changes into gcc's C front-end. Due to this
limitation, C programs that rely on the new instructions have to use inline
assembly to access posit functionality on Clarinet. Further, since the posit data
type is not recognized by the C compiler, direct use of the \texttt{PLW} and
\texttt{PSW} instructions are not possible from C code. In order to circumvent
these limitations we have experimented with different approaches to work with
real-number data as described in section~\ref{sec:clarinet-eg}. 

\subsection{Hardware considerations to integrate the quire}
\label{sec:intgQuire}
As indicated in Fig.~\ref{fig:quireexample} the recommended size
of the quire can grow very rapidly with increasing posit-width.
This implies that treating the quire register similar to an entry
in one of the register files would be quite expensive as far as
hardware resources are concerned. For instance, using 32-bit
posits would mean making a 512-bit quire value available on the
forwarding paths and from the register files. Further, providing a
path from quire to memory (via modified load and store
instructions) would require extensive modifications of the memory
pipeline. Clarinet takes an alternate approach to integrating the quire --
treating it as an accumulator. Quire values can only be manipulated indirectly
using the \texttt{FCVT.R.P} and \texttt{FCVT.P.R} to initialize and read the
quire register respectively. Additionally, the four compute instructions:
\texttt{FMA.P}, \texttt{FMS.P}, \texttt{FDA.P}, and \texttt{FDS.P}, compute
results into the quire register. These decisions allow us to contain the cost
of integrating the quire to just the actual storage for the quire register.
Section~\ref{sec:quire} discusses the design challenges in designing the quire
datapath.

%% file: melodica.tex
Melodica is a posit arithmetic unit implemented using BSV HL-HDL.
Melodica is configured using three parameters: the posit-width (\texttt{N}), the maximum width of the exponent field (\texttt{es}), and float-width.
While Melodica supports any size float input but for Clarinet float-width has been set to 32.
In Clarinet we have experimented with (8,0), (16,1) and (32,2) configurations for the \texttt{N} and \texttt{es} fields.
For an \texttt{N}-bit Melodica architecture $\frac{N^2}{2}$ sized quire is integrated with the operation pipelines as an accumulator register.
Depending on the size of \texttt{N}, it is possible that the quire may not be sized to a multiple of byte.
All fused operations accumulate into the quire.
The four supported fused operators can also be used for basic arithmetic operations.
In addition to the fused operations Melodica implements a set of converters between floating-point and posit formats, and operations to initialize the quire from a poist value and to read the quire as a poist value.

\begin{figure}[!t] 
\centering
\subfloat[Melodica block diagram]{
\includegraphics[width=0.60\columnwidth]{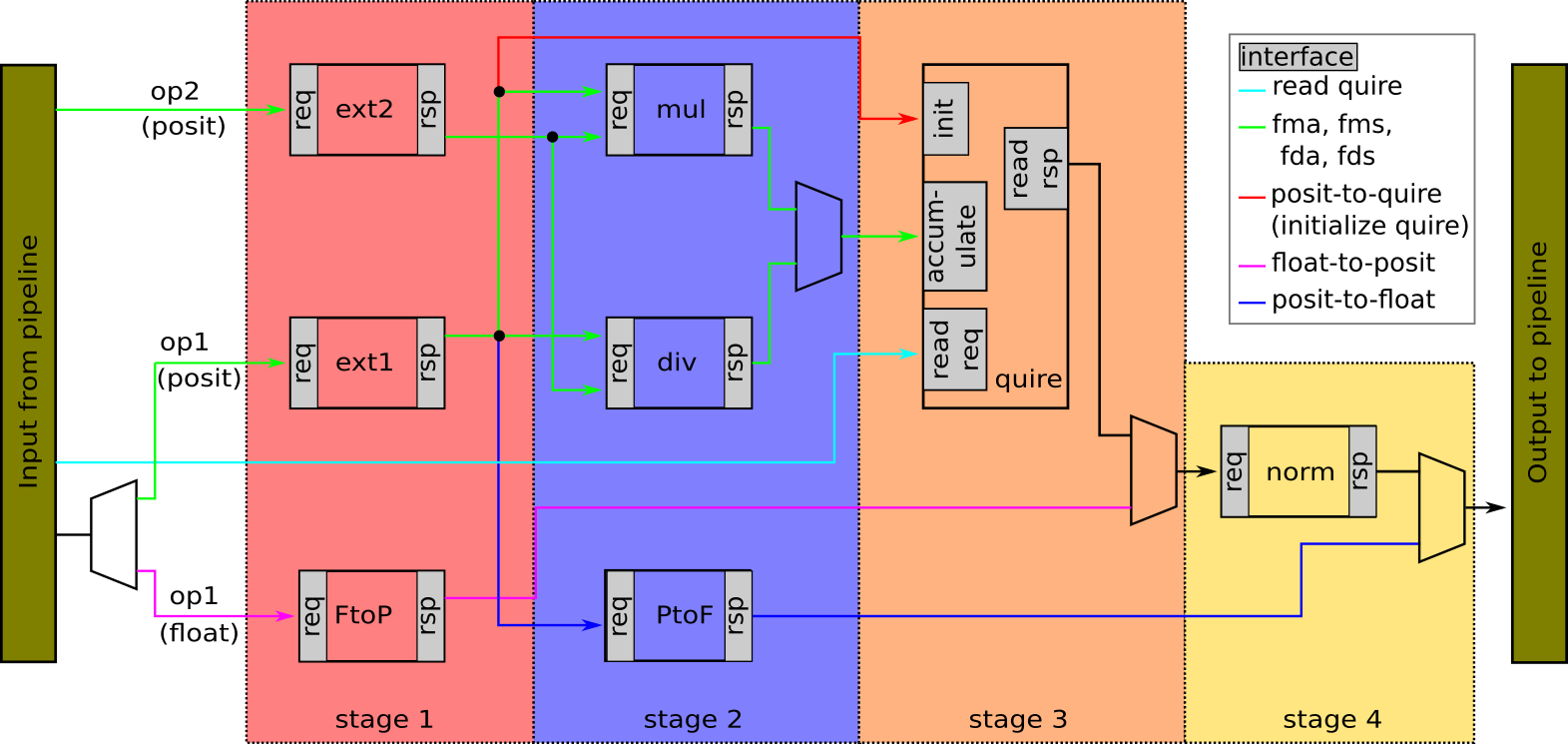}
\label{fig:melodica}}
\subfloat[Quire internals]{
\includegraphics[width=0.35\columnwidth]{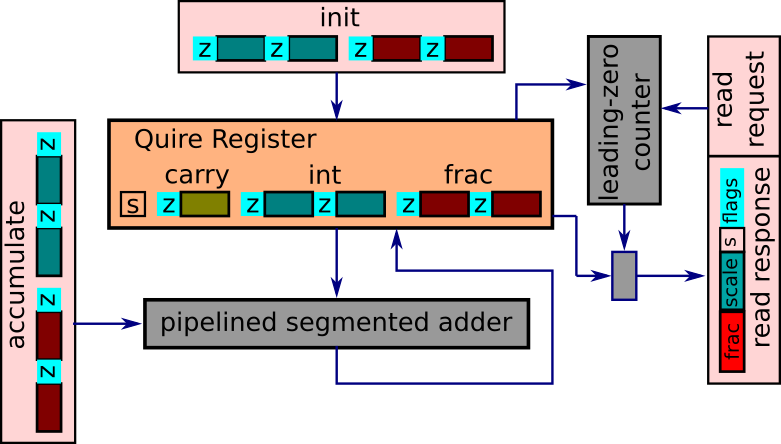}
\label{fig:quire}}
\caption{Melodica block diagram and Quire internals}
\vspace{-5mm}
\end{figure}

Melodica's organization is illustrated in Fig.~\ref{fig:melodica}.
Meodica's input from the pipeline comprises upto two operands, and an opcode.
The CPU pipline issues a command to Melodica only when the pipline is executing one of the special instructions described in section~\ref{sec:custom-instr} that operate on posits.
The Table~\ref{tab:instn-mel-command-mapping} maps Clarinet special instructions to Melodica commands. There are four stages involved in Melodica's operation:

\begin{table}[!b]
\tiny
\centering
\caption{Mapping Clarinet instructions to Melodica commands}
\begin{tabular}{|l|l|l|l|l|p{0.5cm}|p{4.0cm}|}
\hline
\multirow{2}{*}{Clarinet Instruction} & \multicolumn{3}{c|}{Melodica Input} & \multirow{2}{*}{Melodica Output} & \multirow{2}{*}{Stages} & \multirow{2}{*}{Pipelines} \\
\cline{2-4}
         & Operand 1 & Operand 2 & Command   &        &  & \\
\hline
FMA.P    & Posit     & Posit     & FMA\_P    & \xmark & 3 & ext1, ext2, multiply, quire.accumulate \\
FMS.P    & Posit     & Posit     & FMS\_P    & \xmark & 3 & ext1, ext2, multiply, quire.accumulate \\
FDA.P    & Posit     & Posit     & FDA\_P    & \xmark & 3 & ext1, ext2, divide, quire.accumulate \\
FDS.P    & Posit     & Posit     & FDS\_P    & \xmark & 3 & ext1, ext2, divide, quire.accumulate \\
FCVT.R.P & Posit     & \xmark    & FCVT\_R\_P& \xmark & 2 & ext1, quire.init \\
FCVT.P.R & \xmark    & \xmark    & FCVT\_P\_R& Posit  & 2 & quire.read, norm \\
FCVT.P.S & Float     & \xmark    & FCVT\_P\_S& Posit  & 2 & FtoP, norm \\
FCVT.S.P & Posit     & \xmark    & FCVT\_S\_P& Float  & 2 & ext1, PtoF \\
\hline
\end{tabular}
\label{tab:instn-mel-command-mapping}
\end{table}

\begin{enumerate}
   \item \textbf{Extraction:}: posit operands have variable length fields.
      In the extraction stage the fields in the posit operands are decoded by the two extractor blocks (\texttt{ext1} and \texttt{ext2}), and broken up into their component sign, regime, exponent and fraction fields.
      This stage also classifies if the posit operands represent special cases (zero or infinity).
      If the input operand is a float (float-to-posit operations), no extraction is required and the operand is forwarded to the request interface of the \texttt{FtoP} block.
   \item \textbf{Computation:} performs the appropriate (multiply or divide) operation in the case of a fused command from the pipeline (\texttt{FMA.P, FMS.P, FDA.P, FDS.P}.
      The input to this stage is the output of the extractor blocks in the previous stage.
      If the command is \texttt{FCVT\_S\_P} (posit to float conversion), the output of the extractor from the previous stage is fed to the \texttt{PtoF} block.
   \item \textbf{Quire:} involves the \texttt{quire} module and depending on the command being executed by Melodica, different commands are issued to the \texttt{quire} as listed in Table~\ref{tab:instn-mel-command-mapping}. For \texttt{FMA.P, FMS.P, FDA.P, FDS.P} commands, the output of the multiply or divide operations in the previous stage connect into the quire using its \texttt{accumulate} interface.
      If the command is \texttt{FCVT\_R\_P} (initialize the quire), the output of \texttt{ext1} in stage one is used to initialize the quire using its \texttt{init} interface.
      For commands to read the quire (\texttt{FCVT\_P\_R}), the read request is presented to the quire at its \texttt{read.request} interface in this stage as well.
   \item \textbf{Normalization:} is the final stage where the output is prepared for the CPU pipeline.
      The output from the \texttt{read.response} interface of the quire is used as input to the normalizer. The normalizer may also receive input from the \texttt{FtoP} output of stage 1. If the output is of type float (output of \texttt{PtoF}), it is directly multiplexed with the output of this stage before forwarding to the CPU.
\end{enumerate}

\subsection{The Quire Module} \label{sec:quire}
The \texttt{quire} module implements the fixed-point $\frac{N^2}{2}$-bit wide accumulator in Melodica. Three operations are permitted on the quire: initialization, accumulation and read-out (as posit). A block diagram of the quire's internals is illustrated in Fig~\ref{fig:quire}.

The width of the quire has a quadratic relationship with posit width. Consequently in order to scale to wider posits, it becomes necessary to segment the quire and pipeline the accumulate operation~\cite{kulisch1}. In this implementation the segment size (which is a high-level parameter) has been fixed at $32$. This number was chosen as the 32-bit segment adders comprising the pipelined adder would be no wider than the integer adders already in the Clarinet processor core. A side-effect of this choice is that when operating with 8-bit posits, accumulates complete in a single cycle. A \texttt{zero} flag accompanies each segment of the quire, and denotes if the segment is zero or not. The \texttt{zero} flag facilitates a fast (single cycle, fixed latency) count of the leading zeros in the quire which is needed to complete the read operation that converts the fixed-point value to fraction and scale fields for the normalization stage.


%% file: case.tex
We cover case studies using some of the linear algebra kernels and from optical
flow in computer vision. Firstly, we discuss application building in the Clarinet
framework using both, assembly programming and later with C programs. Secondly, we
look into application kernels that are rich in floating-point arithmetic
operations - matrix operations and optical flow. For matrix operations, we develop
a subset of basic linear algebra subprograms (BLAS) and linear algebra packages
(LAPACK) using SoftPosit calls for the error analyses. Based on this
investigation, we arrive at a suitable arithmetic size for each of the kernels in
BLAS and LAPACK, and optical flow estimation using Lucas-Kanade method. We use
this information to set HDL parameters to arrive at customized Melodica and
Clarinet instances. Finally, we execute these applications on the Clarinet CPU
using RTL simulations and report number of cycles.

Table~\ref{tab:experimentalsetup} summarizes the scope of empirical studies using
BLAS, LAPACK, and optical flow as applications. The applications were executed
using SoftPosit calls for error analysis and later simulated on Clarinet RTL for
latency analysis. In addition all hardware configurations used in the study were
synthesized for ASIC and FPGA. The RTL simulation infrastructure, modified RISC-V
compiler toolchain, HDL source code for Melodica and Clarinet along with reference
SoCs and applications are available at~\cite{melodicagit} and ~\cite{clarinetgit}.

\begin{table}[!h]
\tiny
\centering
\caption{Empirical Studies on Clarinet. NR = Not Required.}
\begin{tabular}{|p{2.0cm}| P{0.3cm} | P{0.3cm}| P{0.3cm}| P{0.5cm}| P{0.5cm}| P{0.8cm}| P{0.8cm}| P{0.8cm}| P{0.3cm}| P{0.3cm}|}
\hline
Data types & p8 & p16  & p24 & p32  & p8-q8 & p16-q16 & p24-q24 & p32-q32 & f32 & f64 \\
\hline
SoftPosit \& SoftFloat       & \cmark & \cmark & \cmark & \cmark & \cmark & \cmark & \cmark & \cmark & \cmark & \cmark  \\
Clarinet (synthesis)        & \NR & \NR & \NR & \NR & \cmark & \cmark & \cmark & \cmark & \cmark & \cmark \\
Clarinet (latency)        & \NR & \NR & \NR & \NR & \cmark & \cmark & \cmark & \cmark & \cmark & \cmark \\
\hline
\end{tabular}
 \label{tab:experimentalsetup}
\end{table}

\subsection{Building Posit applications using Clarinet} \label{sec:clarinet-eg}

We begin by illustrating how a programmer can create applications
which use posits in Clarinet. We look at two
scenarios -- an application where the operand data is represented
as Posits in the memory (typically a new application), and a
legacy application which operates on float-point operands where a
developer might want to introduce posits for their computational
advantages.

\subsubsection{Vector dot product}
We illustrate how to program Clarinet to implement a simple vector dot product loop using two assembly code snippets.
In the assembly code samples in~\ref{code:vdp-asm-f} and~\ref{code:vdp-asm-p}, register names starting with \texttt{p} belong to the PRF, while register names starting with \texttt{x} belong to the GPR.
Floating-point registers from the FPR have names starting with \texttt{f}.

  \begin{minipage}[b]{0.48\textwidth}
\begin{Verbatim}
; ------------------------------------------------
; x1, x2 : Vectors A and B base addresses
; x3     : Vector length 
; q      : the quire register
; f2p,p2f: float-to-posit, posit-to-float
                                                
initq: fcvt.r.p p0         ; initialize quire 
                                                
loop:  flw      f1, 0 (x1) ; f1 <- load float A
       flw      f2, 0 (x2) ; f1 <- load float B
       fcvt.p.s p1, f1     ; p1 <- f2p (f1)    
       fcvt.p.s p2, f2     ; p2 <- f2p (f2)    
       fma.p    p1, p2     ; q <- q + (p1 * p2)
       add      x1, x1, 4  ; increment A index 
       add      x2, x2, 4  ; increment B index 
       sub      x3, x3, 1  ; decrement loop ctr
       bnez     x3, loop                       
                                               
readq: fcvt.p.r p0         ; p0 <- read (q)    
       fcvt.s.p f0, p0     ; f0 <- p2f (p0) 
\end{Verbatim}
  \label{code:vdp-asm-f}
  \end{minipage}
  \quad 
  \begin{minipage}[b]{0.48\textwidth}
\begin{Verbatim}
; ------------------------------------------------
; x1, x2 : Vectors A and B base addresses
; x3 : Vector length
; q  : the quire register

initq: fcvt.r.p p0         ; initialize quire

loop:  plw      p1, 0 (x1) ; p1 <- load posit A
       plw      p2, 0 (x2) ; p2 <- load posit B

       fma.p    p1, p2     ; q <- q + (p1 * p2)
       add      x1, x1, 4  ; increment A index
       add      x2, x2, 4  ; increment B index
       sub      x3, x3, 1  ; decrement loop ctr
       bnez     x3, loop
                                                
readq: fcvt.p.r p0         ; p0 <- read (q)

\end{Verbatim}
  \label{code:vdp-asm-p}
  \end{minipage}
  
The code sample on the right in~\ref{code:vdp-asm-p} illustrates the core loop of the vector dot product computation where the operands are represented as posits in memory.  
Due to the quire-based computation model, any computation with posits in Clarinet involves three steps: i) initializing the quire, ii) computing into the quire using one or more of the four compute instructions, and iii) reading the quire.
Initializing and reading the quire represent a fixed overhead of two instructions.
Since these do not form part of the core computation loop, their effect can be ignored for loops with a large number of iterations, but may become substantial when these operations are frequent or for loops with few iterations as illustrated in~\ref{sec:blaslapack}.

The code sample on the left in~\ref{code:vdp-asm-f} illustrates the same dot-product program with float operands. The data being operated on is stored as floats in memory however the computations continues to use posits.
The three steps involved in quire-based computation remain unchanged.
However, the computation step involves first converting the float operands into posits.
This leads to an overhead of two instruction per iteration of the compute loop, and represents a significant overhead compared to pure float (or posit) based computation flow.
The mixed mode of computation may find favour in legacy implementations of applications where a transition of the working data from floats to posits may prove problemmatic.
In such cases, the mixed mode of computation would allow the application developer to insert posit based computation where necessary into their algorithm to take advantages of greater accuracy or dynamic range while retaining the working data in its original format.

\subsubsection{Abstracting away from assembly instructions} For the larger BLAS,
LAPACK and optical flow examples, we packaged the three steps involved in
quire-based computation into C functions using inline assembly for ease of use and
debug. The functions in~\ref{code:c-fns} are examples of such helper functions.
Since our present work does not include supporting the posit data type as a
primitive data type in C, the compilers do not recognize the \texttt{posit} type.
In order to circumvent this limitation we reference posit values as
\texttt{unsigned char} (8-bit posits) or \texttt{unsigned short} (16-bit posits)
or \texttt{unsigned int} (24 and 32-bit posits). These variables are read from or
written to memory using integer load and store instructions, followed by the
\texttt{pmv} instructions to move the posit values into the PRF for further
processing.

  \begin{minipage}[b]{0.48\textwidth}
\begin{Verbatim}
// ------------------------------------------------
// Initialize the quire, given a 32-bit posit
// initVal (stored as unsigned int in memory)
void fn_init_p_quire (unsigned int initVal) {

   // Move the GPR value to the PPR
   volatile asm ("pmv.w.x fa4, %[inA]\n"
                 "fcvt.r.p ft0, fa4"
                 : : [inA] "r" (initVal) : "fa4");
   return;
}
\end{Verbatim}
  \label{code:c-fns-init-q}
  \end{minipage}
  \quad 
  \begin{minipage}[b]{0.48\textwidth}
\begin{Verbatim}
// ------------------------------------------------
// Read the quire into a posit register and return
// the value as unsigned int
unsigned int fn_read_p_quire (void) {
   unsigned int res;
   // Move the posit value from PPR to GPR.
   asm volatile("fcvt.p.r ft0, ft0 \n"
                "pmv.x.w  %[o_res], ft0" 
                : [o_res] "=r" (res) : : "ft0");
   return (res);
}
\end{Verbatim}
  \label{code:c-fns}
  \end{minipage}
  
  \begin{minipage}[b]{0.48\textwidth}
\begin{Verbatim}
// ------------------------------------------------
// Perform a fused multiply accumulate into the
// quire with two 32-bit posit values as inputs
// (the posits are represented as unsigned int)
void fn_positFMA (
        unsigned int a, unsigned int b) {

   // Move input values from GPR into the PPR
   asm volatile (
         "pmv.w.x fa4, %[inA]\n"
         "pmv.w.x fa5, %[inB]\n"
         "fma.p   ft0, fa4, fa5"
         : : [inA] "r" (a), [inB] "r" (b)
         : "ft0", "fa4", "fa5");    // clobber

   return;
}
\end{Verbatim}
  \label{code:c-fns-fma}
  \end{minipage}
  \quad 
  \begin{minipage}[b]{0.48\textwidth}
\begin{Verbatim}
// ------------------------------------------------
// Dot product of two vectors of 32-bit posits. The
// arrays are stored as unsigned int in memory
unsigned int fn_positVDP (int r,    // array size
   unsigned int p_a[], unsigned int p_b[]) {

   unsigned int acc = 0;   // posit-32 output
   int i = 0;
   fn_init_p_quire (acc);  // quire <- 0

   // The dot product loop, using quire
   for (i=0; i<r; i++) fn_positFMA (p_a[i], p_b[i]);

   // Read result from quire as posit-32
   acc = fn_read_p_quire ();
   return (acc);
}
\end{Verbatim}
  \label{code:c-fns-vdp}
  \end{minipage}


\subsection{BLAS and LAPACK}\label{sec:blaslapack}

\subsubsection{Error Analysis using SoftPosit} \label{sec:soft-error-analysis}

\begin{figure}[!t]
    \includegraphics[width=\columnwidth]{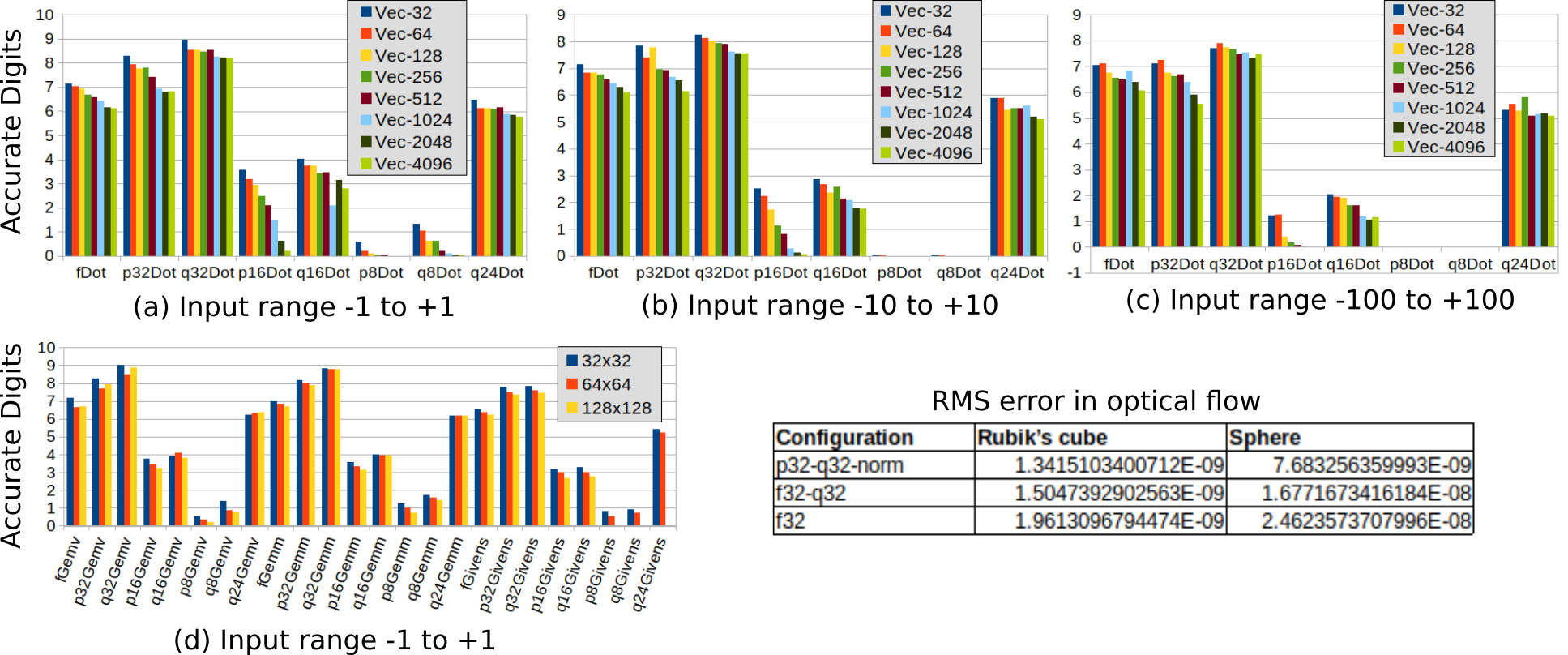}
    \caption{Accurate number of digits with different data types in BLAS and LAPACK routines, and RMS error in optical flow}
    \vspace{-5mm}
    \label{fig:blasplots}
\end{figure}
The BLAS and LAPACK are encountered in a wide range of engineering and scientific applications. In BLAS, we consider dot product (\texttt{xDot}), matrix-vector (\texttt{xGemv}), matrix-matrix operations (\texttt{xGemm}), and in LAPACK, we consider Givens rotation (\texttt{xGivens}) where \texttt{x} denotes the data type used for the implementations. 
For all the matrix operations, we implement nine different versions using different data types for comparison and use 64-bit floating-point implementation as a reference. We randomly generate numbers using \texttt{rand()} function. Since, Clarinet supports quire and FMA, we emphasize more on quire based implementations with using SoftPosit for our analyses. To calculate the error in \texttt{xDot}, we average the relative error over 100K runs. To calculate error in \texttt{xGemv} and \texttt{xGemm}, we use $\frac{\parallel \hat{x}-x \parallel _2}{\parallel x \parallel _2}$ and $\frac{\parallel \hat{A}-A \parallel _2}{\parallel x \parallel _2}$, respectively where $x$ and $A$ are the operations computed in 64-bit floating-point and $\hat{x}$ and $\hat{A}$ are the operations computed using SoftPosit.  

The accurate digits in the different implementations are shown in Fig.~\ref{fig:blasplots}. In dot product, the 32-bit quire (\texttt{q32Dot}) results in 8.8 accurate digits for small ($<$10) input vector sizes in range of 0 to 1 (Fig. \ref{fig:blasplots}a). For large vectors ($<$10000) in the same range, the number of accurate digits drop to 8.2 which is a drop of 6.8\%. In the same input rage, we observe a drop of 12.3\% in \texttt{fDot}, 17.4\% in \texttt{p32Dot}, and 9.3\% in \texttt{q24Dot}. For the input vector range of 0 to 10 and the sizes of 10 to 10000, we observe a similar trend (Fig. \ref{fig:blasplots}b). Varying the range of input vectors impacts the accuracy heavily, specifically for large vectors. We observe a drop in the number of accurate digits by 55.6\% in \texttt{q32Dot}, 53.94\% in \texttt{p32Dot}, and 36.3\% in \texttt{q24Dot} while in \texttt{fDot} it is 18.63\% (Fig.~\ref{fig:blasplots}c). The drop in accuracy is due to the fact that the posit and quire are more accurate for the values around 1.0 while as the input range shifts from 1.0 the accuracy deteriorates. 

A similar trend is observed in \texttt{xGemv}, \texttt{xGemm}, and \texttt{xGivens} routines for increasing matrix sizes and varying ranges (Fig.~\ref{fig:blasplots}d). A key observation here is that in \texttt{p32Givens} and \texttt{q32Givens} routines where we observe  the number of accurate digits are significantly higher (8.2 and 8.8 respectively) compared to \texttt{fGivens} (6.79). The shaded region in Fig.~\ref{fig:blasplots} represents the routines that can be executed on the current version of Clarinet due to absence of posit addition, multiplication, division hardware. For implementation of routines in software we have used floating-point in conjunction with quire. For example, \texttt{q32-f32Givens} is implementation of Givens rotation using combination of 32-bit quire and 32-bit floating-point arithmetic. The implementation yields similar accuracy as \texttt{q32Givens} since the majority of the operations are dominated by quire. In BLAS routines, the 100\% of the arithmetic operations can be implemented using only quire. 

\subsubsection{Simulation on Clarinet RTL}
The intent of simulating applications on Clarinet RTL was to study the effects that different real number types, namely floating-point and posits had on latency of computation, and also to come with best practices to use posit hardware more effectively in a real RISC-V CPU. Our test rig consisted of RTL simulators for the different RTL configurations tabulated in table~\ref{tab:clarinet_configs}, elf files generated by compiling the applications and kernels using the modified RISC-V compiler described in section~\ref{sec:compiler}, and tools to extract timing information from instruction traces generated during RTL simulations.

Applications for Clarinet can adopt any one of three approaches with respect to real-number types: (i) single or double-precision floating point where the data in memory resides as floats and the computation is carried out using floating-point arithmetic pipelines, (ii) float-posit where the data in memory resides as floats but the computation is carried out using posit arithmetic pipelines, and (iii) posit where the data in memory resides as posits and the compution is carried out using posit arithmetic pipelines. We implemented the BLAS kernels and LAPACK routines using all three approaches, and across four different posit widths -- 8, 16, 24 and 32.  These implementation configurations are summarized in the
Table~\ref{tab:xDot_variations}.

\begin{table}[!b]
\tiny
\centering
\caption{Implementation Variations (wrt real number types)}
\begin{tabular}{|p{1.5cm}| p{2.3cm} | p{2.3cm}| p{3.4cm}| P{1.0cm}| P{1.0cm}|}
\hline
Configuration & Input Data                & Output Result            & Computation                                & FPR-PRF Conversion & GPR-PRF Move          \\
\hline                                                                                                               
f32           & FPR loads (flw)           & FPR stores (fsw)         & FP fused multiply-add (fmadd.s)            & \xmark & \xmark                \\
f32-p8        & FPR loads (flw)           & FPR stores (fsw)         & 8-bit Quire fused multiply-add (fma.p)     & \cmark & \xmark                \\
f32-p16       & FPR loads (flw)           & FPR stores (fsw)         & 16-bit Quire fused multiply-add (fma.p)    & \cmark & \xmark                \\
f32-p32       & FPR loads (flw)           & FPR stores (fsw)         & 32-bit Quire fused multiply-add (fma.p)    & \cmark & \xmark                \\
p8            & GPR byte loads (lbu)      & GPR byte stores (sbu)    & 8-bit Quire fused multiply-add (fma.p)     & \xmark & \cmark                \\
p16           & GPR half-word loads (lhu) & GPR half-word stores (shu) & 16-bit Quire fused multiply-add (fma.p)  & \xmark & \cmark               \\
p32           & GPR word loads (lw)       & GPR word stores (sw)     & 32-bit Quire fused multiply-add (fma.p)    & \xmark & \cmark                  \\
\hline
\end{tabular}
\label{tab:xDot_variations}
\end{table}

\begin{figure}[!t]
\centering
\subfloat[BLAS Cycles (absolute)]{
    \includegraphics[width=0.48\columnwidth]{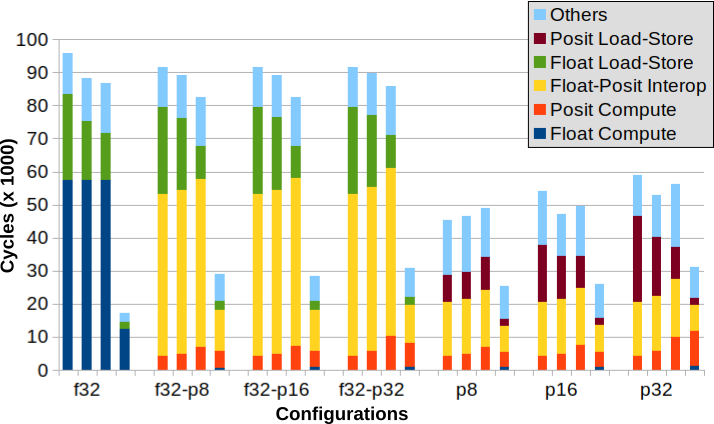} \label{fig:blas-cyc}}
\subfloat[BLAS Cycles (\%)]{
    \includegraphics[width=0.48\columnwidth]{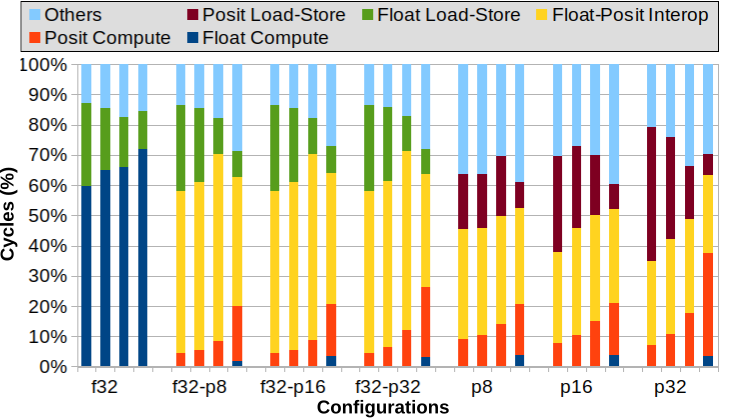} \label{fig:blas-percent}}
\caption{BLAS kernels (xDot, xGemv, xGemm) and xGivens cycle counts}
\vspace{-5mm}
\end{figure}
\begin{figure}[!t]
\centering
\subfloat[BLAS Cycles (absolute)]{
    \includegraphics[width=0.48\columnwidth]{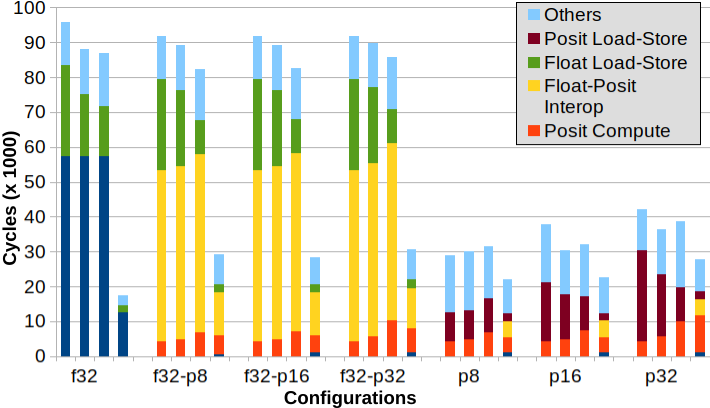} \label{fig:blas-ideal-cyc}}
\subfloat[BLAS Cycles (\%)]{
    \includegraphics[width=0.48\columnwidth]{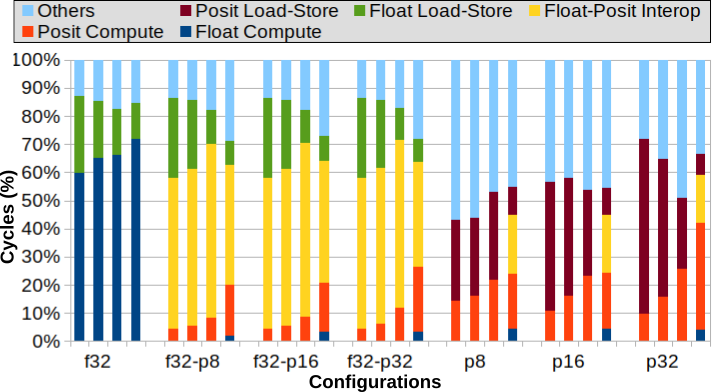} \label{fig:blas-ideal-percent}}
\caption{BLAS kernels (xDot, xGemv, xGemm) and xGivens cycle counts (estimated with posits as a primitive data type)}
\vspace{-5mm}
\end{figure}

Fig~\ref{fig:blas-cyc} represents the number of clock cycles consumed by different groups of instructions while executing the \texttt{xDOT}, \texttt{xGEMV}, \texttt{xGEMM}, and \texttt{xGivens} kernels. Fig~\ref{fig:blas-percent} breaks up the same information into percentages of the total clocks cycles consumed to execute the kernels. In order to present data across the three BLAS kernels, the vector and matrix sizes input to the kernels were chosen to require the same number of floating-point or posit (multiply-accumulate) operations. Consequently, \texttt{xDOT} uses 4096 element vectors, \texttt{xGEMV} uses 64 element vectors and 64x64 matrices, and \texttt{xGEMM} uses 16x16 matrices. Normalizing across number of computations however implies that the amount of data required for each kernel reduces with the increasing order of the kernel. All kernels were run with warm caches and the \texttt{-O3} optimization flag was used to compile the C code. In the case of \texttt{xGivens}, the matrix size was reduced to 8x8. The cycles required to execute \texttt{xGivens} on a 16x16 matrix was more than double the cycles required for the BLAS kernels - halving the size allowed us to fit the data from \texttt{xGivens} in the same scale. For the analysis, instructions were grouped as per the following categories:
\begin{enumerate}
   \item \textbf{float compute} instructions directly involved with floating-point computation. In the case of the BLAS kernels these are \texttt{fmadd.s}, \texttt{fmul.s}, and \texttt{fadd.s}. In other applications like givens, this category would also include other floating-point instructions like \texttt{fsqrt.s} and \texttt{fdiv.s} 

   \item \textbf{float ld/st} instructions to load and store floating-point data
      from and to memory. For single-precision operation these are the
      \texttt{flw} and \texttt{fsw} instructions.

   \item \textbf{posit compute} instructions that perform computations into the quire. These are \texttt{fma.p}, \texttt{fms.p}, \texttt{fda.p}, and \texttt{fds.p}. Instructions to initialize the quire (\texttt{fcvt.r.p}) and read the final result from the quire (\texttt{fcvt.p.r}) are also included in this category as initializing the quire and reading out its value as a posit value form the bookends of any posit computation in Clarinet.
      
   \item \textbf{float-posit interop} instructions that are present to allow
      interoperability between floating-point and posit values. These include instructions like \texttt{fcvt.s.p} and \texttt{fcvt.p.s} to convert between floating-point and posit values. This category also includes the instructions to move posit values between the GPR and PRF (\texttt{pmv.x.w} and \texttt{pmv.w.x}). The \texttt{pmv} instructions are not used strictly for interoperability between floats and posits, however their presence is needed because of the lack of high-level compiler support for the posit data type.

   \item \textbf{posit ld/st} instructions to load and store posit data from and to memory. Due to the lack of higher-level compiler support for the posit data type, these appear as 8, 16 and 32 bit unsigned loads and stores in the trace (\texttt{lbu} and \texttt{sbu}, \texttt{lhu} and \texttt{shu}, and \texttt{lw} and \texttt{sw}).

   \item \textbf{others} all other instructions in the trace not categorized as one of the above.
\end{enumerate}

The absolute number of cycles spent on posit compute in the BLAS kernels is consistently lower than those spent on float compute. The main reason for more efficient posit compute is better use of pipelining in the posit core. As instructions that update the quire do not have a side-effect on the Clarinet's register files, the posit core can queue up several such instructions leading to more effective use of the deep posit pipeline. Compared to an individual \texttt{FMADD.S} instruction that takes 12 cycles, a \texttt{FMA.P} instruction may take 12, 20, 36 cycles for 8-bit, 16-bit and 32-bit posits respectively. However, unlike a \texttt{FMADD.S} that responds with the answer after 14 cycles, a \texttt{FMA.P} instruction responds immediately since the answer is going to accumulate into the quire and the pipeline can proceed. Subsequent \texttt{FMA.P} instructions queue into the pipeline, effectively hiding the long latencies associated with the deeper posit pipelines. The concealment of the posit fused operation latency is more effective for narrower posits as these take fewer cycles to compute the result into the quire and subsequently require fewer operations to be queued in to optimally use the pipeline. Since the operating frequency of the FPU and Melodica are very similar for the FPGA devices considered here, the fewer cycles takes for posit compute translate to shorter latencies for posit compute.   

The cycles in yellow in Fig~\ref{fig:blas-cyc} can be considered as an overhead to the use of posits in the clarinet system. When operating in one of the interop modes - floating-point data, posit compute configurations (\texttt{f32-p*}) - this overhead is significantly larger than the posit compute cycles themselves as it includes the cycles spent in converting the floating point data into posit values for each iteration of the computation loop as described in~\ref{sec:clarinet-eg}. These operations are not pipelined as they return a value to the Clarinet's register files. However, when operating with posit data directly (configurations \texttt{p*}), this overhead is smaller (Fig~\ref{fig:blas-percent} and is an artefact of the lack of higher-level compiler support for the posit data type in C. This lack of support requires us to save posit data as unsigned integer types, read/write them via the GPR and then move them over to the PRF to initiate the posit computation. If posit data types were to be treated as a primitive data type in C, the overhead when working with posit data directly would disappear as illustrated in Fig~\ref{fig:blas-ideal-cyc}. However, there is no change seen for the interop modes (\texttt{f32-p*}) as the overhead in these modes are due to switching between floats and posits - this overhead would remain even if posits were to be a primitive data type.

When operating with 8 and 16-bit posit data (\texttt{p8} and \texttt{p16}), memory access latencies are lower due to the more efficient utilization of Clarinet's 8KB level one caches. In the case of \texttt{xDOT} the data set completely fits in the cache only when operating with 8-bit posits, and in all other cases memory latency is dominated by capacity misses. However in higher-order kernels like \texttt{xGEMV} and \texttt{xGEMM} the effect is less pronounced as the data sets are smaller. In the case of \texttt{xGEMM}, the complete data set fits into the cache across all data types. Consequently, the memory accesss latency is near constant across all configurations. For \texttt{xGEMV} the complete data set fits in to the cache when operating with 8 and 16-bit posits. This results in a small advantage in overall cycle counts for 8 and 16-bit posit widths when compared to \texttt{f32}.

Higher-order kernels involve more bookkeeping in the form of cycle counters and function calls to lower-order kernels, and these reflect in the larger number of cycles spent in the unaccounted (others) category. 
When operating with floating-point data (\texttt{f32}) or in one of the interop modes (\texttt{f32-p*}), higher order kernels take fewer number of overall cycles to execute. 
This is due to a greater proportion of the input data fitting into the caches resulting in fewer capacity misses and consequently a lower cycle count for floating point loads and stores. 
Across all configurations that use posits for computation (\texttt{f32-p*} and \texttt{p*}), higher order kernels result in an increase in the cycle spent in computation. 
This effect is especially pronounced in Fig~\ref{fig:blas-ideal-cyc} and Fig~\ref{fig:blas-ideal-percent} due to larger percentage of overall cycles being spent on posit compute. 
While the number of multiply-accumulate operations are constant across all configurations, higher-order kernels require repeated initialization and reading of the quire register (once per call of the \texttt{xDot} kernel). 
In the case of \texttt{xGEMV}, this works out to $O(N)$ quire accesses, and $O(N^2)$ quire accesses for \texttt{xGEMM}.
Requests to initialize the quire complete immediately, as there is no value returned to the Clarinet register files and do not contribute much to these increase in cycles. 
However, reading the quire is an expensive operation from a latency point of view as the pipeline has to wait for all outstanding fused operations to complete before a value can be returned to Clarinet's register files.

\begin{table}[!b]
\tiny
\centering
\caption{Accumulations versus Quire Reads across kernels}
\begin{tabular}{|p{4.0cm}| p{1.5cm} | p{1.5cm}| p{1.5cm}| P{1.5cm}|}
\hline
                                & xDot (4096)  & xGEMV (64)   & xGEMM (16)   & xGivens (8)  \\
\hline                                                                                                               
Quire initializations and reads & 1            & 64           & 256          & 64 \\
Accumulations between quire reads & 4096         & 64           & 16           & 1 \\
\hline
\end{tabular}
\label{tab:accum_depth}
\end{table}

The \texttt{xGivens} application represents a case where all posit operations
cannot be realised in hardware due to limitations in the current Melodica
implementation, namely the unavailability of the square root operator, and interoperability with floats is necessary to complete the   
$O(N^3)$ square root operations in this implementation. 
As Figure~\ref{fig:blas-cyc} indicates, running \texttt{xGivens} on floating-point 
outperforms all posit and float-posit configurations. The main reason for the poorer performance of posit based configurations is the inability to effectively pipeline posit compute operations in this \texttt{xGivens} implementation. Table~\ref{tab:accum_depth} tabulates the number of accumulations between initializing the quire and its subsequent read at the end of a set of computations. The longer the string of uninterrupted accumulations between two quire reads, the more effective is the use of the posit pipeline. Further, narrower posits are more forgiving of shorter accumulation sequences as the pipeline depth for quire accumulation in melodica grows quadratically with posit width (one-stage, four-stage, and 16-stage for 8-bit, 16-bit and 32-bit posits respectively). In the case of the \texttt{xGivens} implementation, the number of accumulations is independent of the matrix size, each posit computation sees an effective latency of 36 cycles for 32-bit posits and the effect is amplified. Even in the case of \texttt{xGEMM} the poor ration of accumulation depth to quire reads leads to a steep increase in the number of computation cycles, especially for 32-bit posits (Fig.~\ref{fig:blas-percent}).

\subsection{Lucas-Kanade optical flow}\label{sec:lk}
\subsubsection{Error Analysis using SoftPosit}
\begin{figure}[!t] 
\centering
\subfloat[Rubik's cube and sphere]{
\includegraphics[width=0.3\columnwidth]{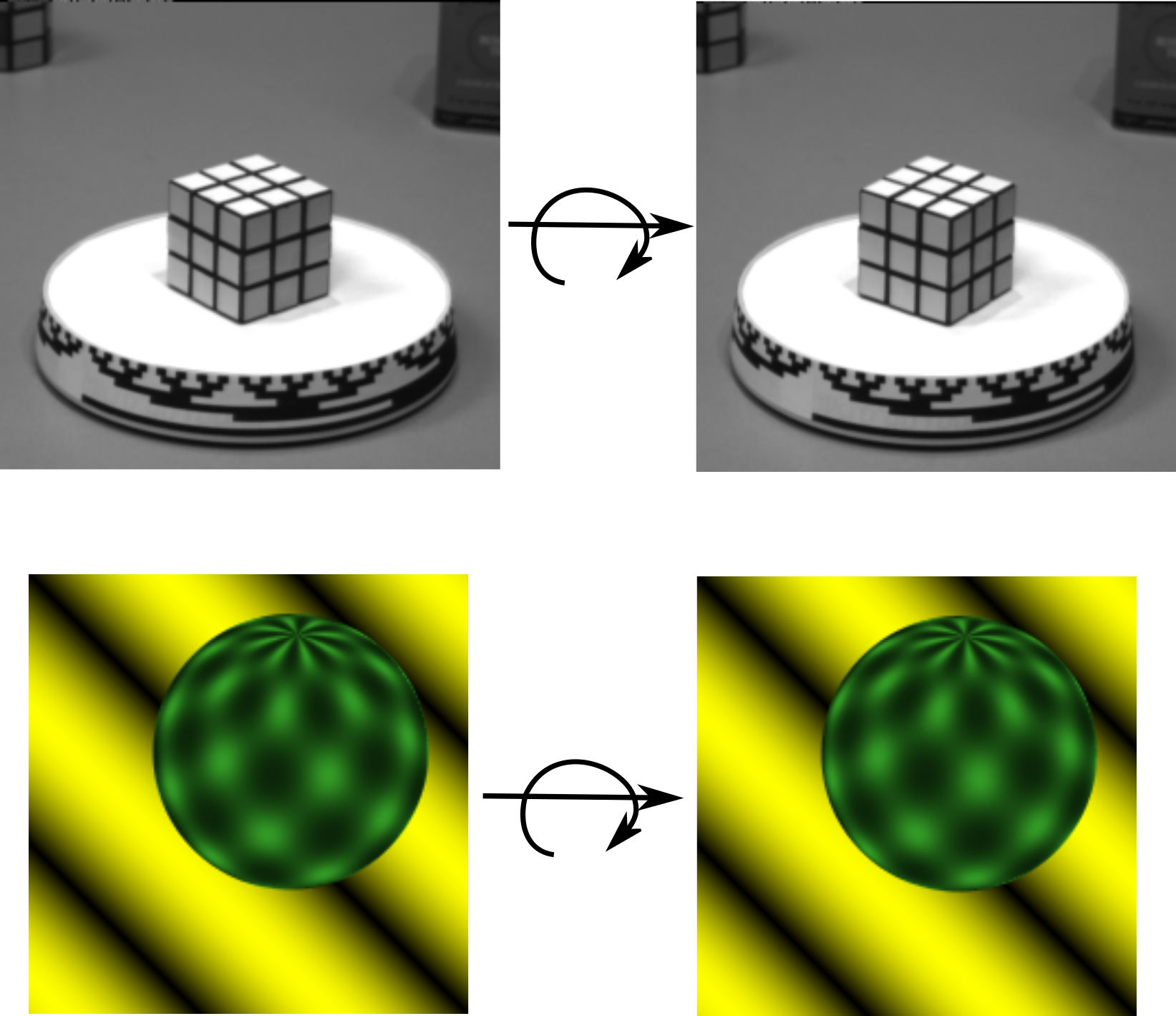}} 
\subfloat[Maximum error and RMS error with respect to 64-bit floating-point]{
\includegraphics[width=0.5\columnwidth]{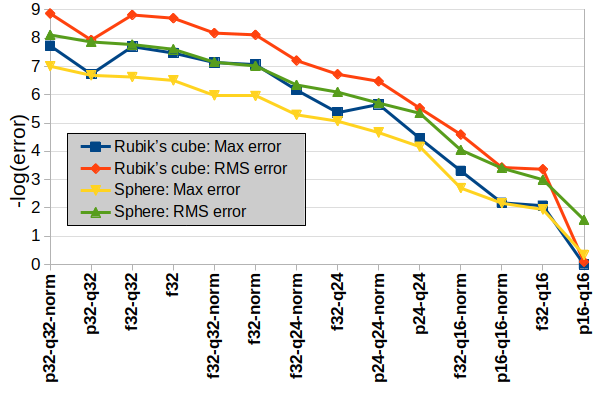}
\label{fig:max_rms}}
\caption{Dataset for Lucas-Kanade and error analysis using SoftPosit}
\vspace{-5mm}
\label{fig:objects}
\end{figure}

Lucas-Kanade is a differential method of tracking features given a sequence of frames. Given \texttt{I} as brightness-per-pixel at \texttt{(x,y)}, the local optical flow (velocity) vector \texttt{($\Vec{u}$,$\Vec{v}$)} is given by $ \frac{\partial I}{\partial x}u + \frac{\partial I}{\partial y}v + \frac{\partial I}{\partial t} = 0$. The Lucas-Kanade method is used to calculate the optical flow for consecutive frames of rotating objects which are given in Fig.~\ref{fig:objects}. We compare the different posit and single-precision floating-point configuration combinations with 64-bit floating-point values using SoftPosits, and generate heat maps of the absolute error for both \texttt{u} and \texttt{v}. The three configurations that are being compared are: i) 32-bit single-precision floating-point arithmetic (\texttt{f32}), ii) 32-bit single-precision float arithmetic combined with \texttt{N}-bit quire arithmetic (\texttt{f32-qN}), iii) \texttt{N}-bit posit arithmetic and N-bit quire arithmetic (\texttt{pN-qN}). Furthermore, owing to the better accuracy of posits around 1.0 we have normalized (\texttt{norm}) grey-scale pixel values (0 to 255) to (0.0 to 16.0).

\begin{figure}[!t]
\subfloat[Error heat-maps of Rubik's cube for pixels between 0-255]{
    \includegraphics[width=0.15\columnwidth]{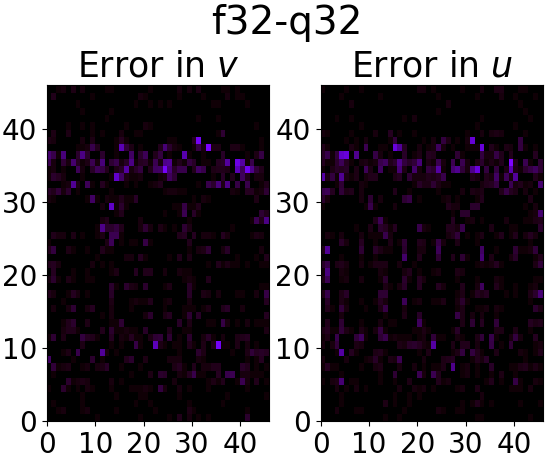}
    \includegraphics[width=0.15\columnwidth]{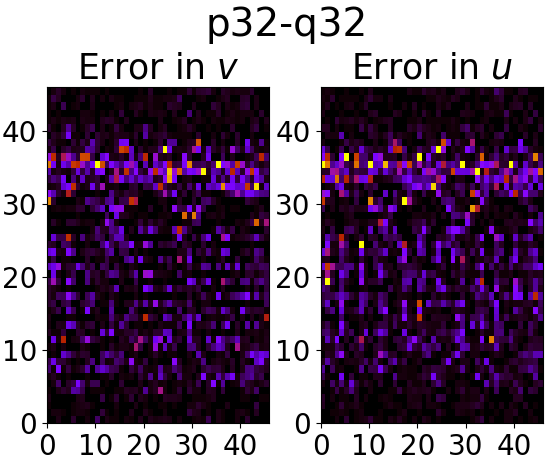}
    \includegraphics[width=0.15\columnwidth]{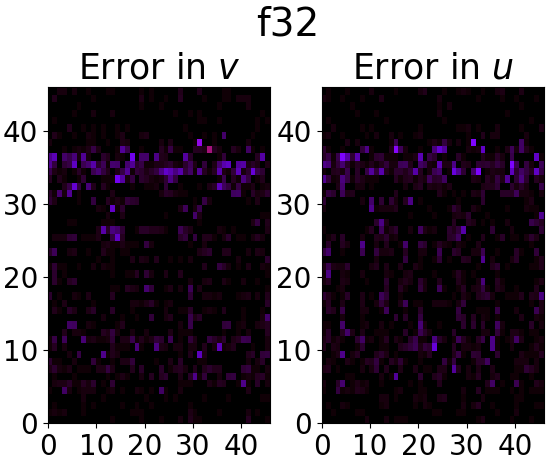}
    \label{fig:rubik_heatmap}}
\subfloat[Error heat-maps of Rubik's cube normalized pixels between 0.0-16.0]{
    \includegraphics[width=0.15\columnwidth]{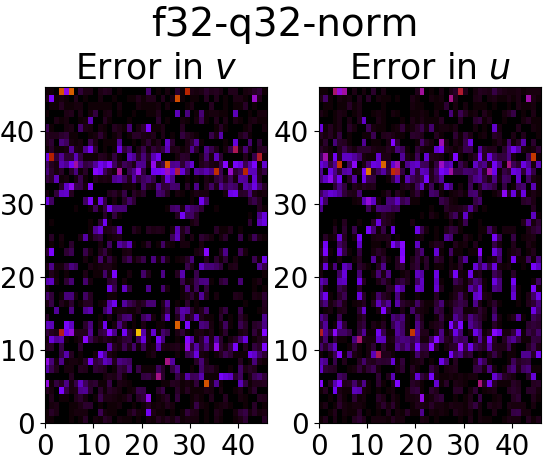}
    \includegraphics[width=0.15\columnwidth]{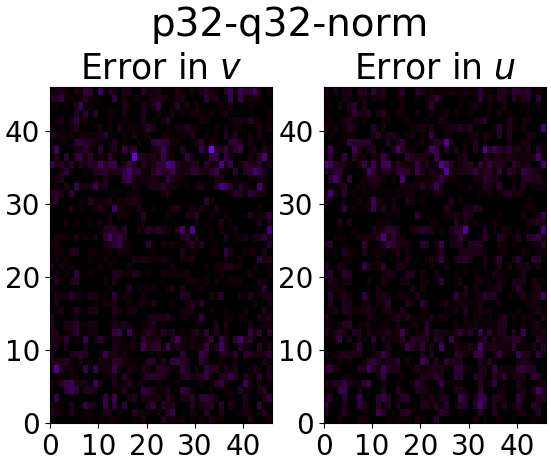}
     \includegraphics[width=0.15\columnwidth]{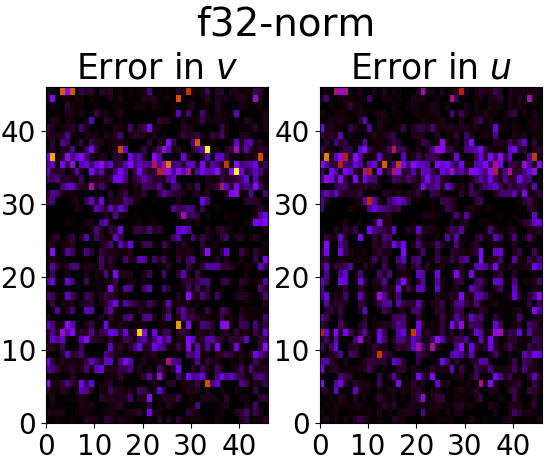}
     \includegraphics[width=0.03\columnwidth]{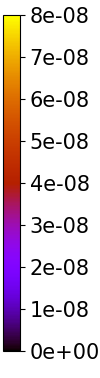}
     \label{fig:rubik_heatmap_n}}
     \\
\subfloat[Error heat-maps of sphere pixels between 0-255]{
    \includegraphics[width=0.15\columnwidth]{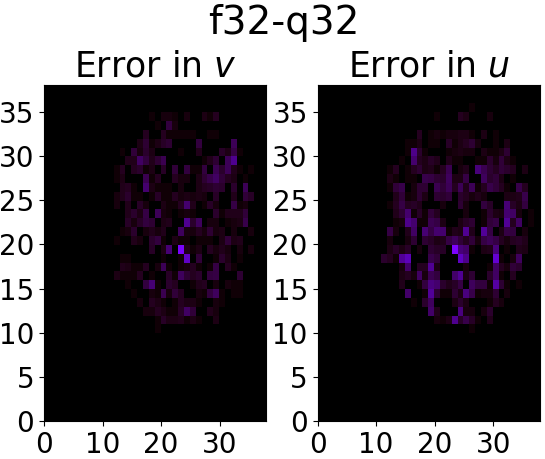}
    \includegraphics[width=0.15\columnwidth]{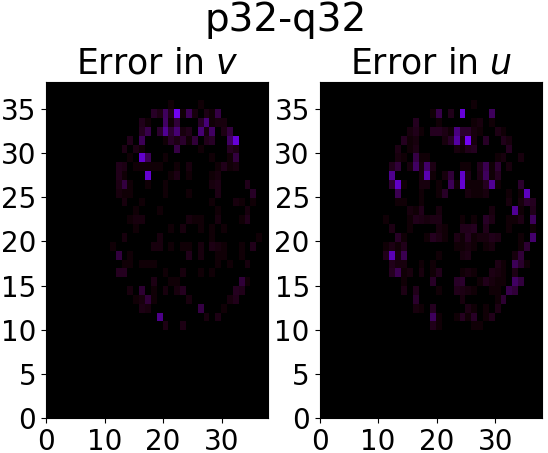}
    \includegraphics[width=0.15\columnwidth]{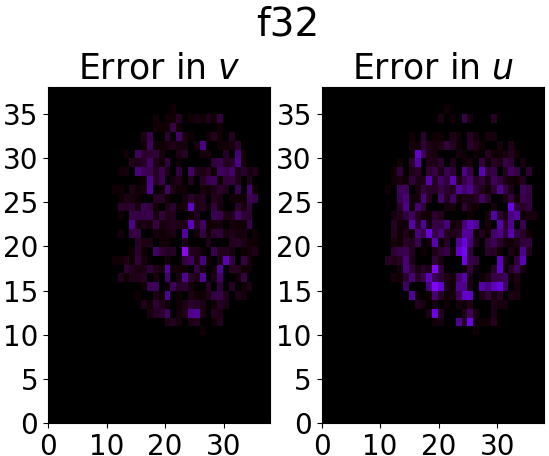}
    \label{fig:sphere_heatmap}}
\subfloat[Error heat-maps of sphere normalized pixels between 0.0-16.0]{
    \includegraphics[width=0.15\columnwidth]{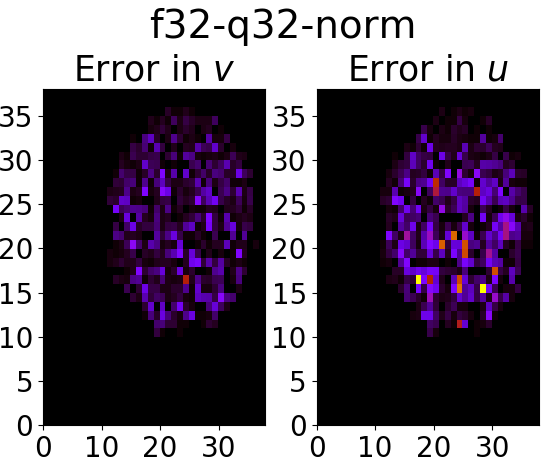}
    \includegraphics[width=0.15\columnwidth]{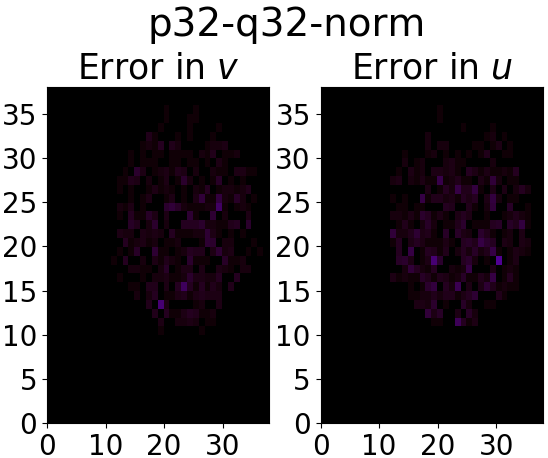}
     \includegraphics[width=0.15\columnwidth]{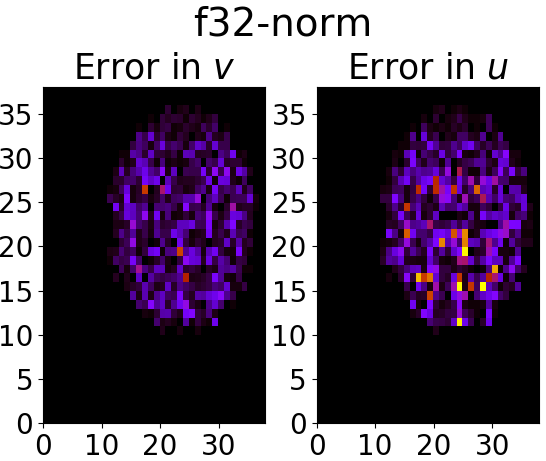}
     \includegraphics[width=0.04\columnwidth]{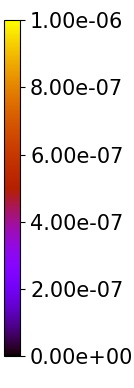}
    \label{fig:sphere_heatmap_n}}
\caption{Error heat-maps for Rubik's cube and sphere}
\vspace{-5mm}
\label{fig:heatmap}
\end{figure}

From the heat-maps in Fig.~\ref{fig:rubik_heatmap} the effects of normalization and \texttt{q32} on error become obvious. When working with normalized data, the configuration \texttt{p32-q32} clearly outperforms all other configurations. For data which is not normalized, the performance of \texttt{p32-q32} depends on whether the data naturally falls around 1.0. However, even in the non normalized case, \texttt{f32-q32} performs consistently better than \texttt{f32}. The general trend in maximum and RMS error for Rubik's cube and sphere object frames for different configurations are shown in Fig.~\ref{fig:max_rms}. The y-axis value for RMS error in Fig.~\ref{fig:max_rms} gives the number accurate digits for the configurations compared to 64-bit floating-point. Allowing one decimal places of tolerance to error, \texttt{p24-q24-norm} configuration can give accurate results close to \texttt{f32}. With a penalty of 2 more decimal place \texttt{p16-q16-norm} can be a feasible alternative. When optical flow is computed for posit  configurations for values not around 1.0 the accuracy falls. As summarised in the satellite Table in Fig.~\ref{fig:blasplots}, \texttt{p32-q32-norm} configuration results in an order improvement in accuracy as compared to \texttt{f32} for the sphere dataset. The \texttt{f32-q32} configuration for gray-scale pixel values (0-255) improves the accuracy by 23\% and 32\% for Rubik's cube and sphere dataset respectively. 
 

\subsubsection{Simulating on Clarinet RTL}
Adapting the Lucas-Kanade optical flow application to run on Clarinet is a good example showcasing the co-existence of floating-point and posit types on Clarinet hardware. In order to take advantage of the quire accumulator in Clarinet only the inner multiply-accumulate loop to compute the velocity is converted to use posits. This corresponds to the \texttt{f32-q32}, \texttt{f32-q24}, and \texttt{f32-q16} cases in the earlier experiments with SoftPosits. 

\begin{minipage}[b]{0.48\textwidth}
\begin{Verbatim}
// Original (with floating point)
float AtA_f[2][2]; int r, c, k;
for (r = 0; r < 2; r++) {
  for (c = 0; c < 2; c++) {
    AtA_f[r][c] = 0.0;
    // accumulation loop (using float in memory)
    for (k = 0; k < Arows; k++) 
       AtA_f[r][c] += (A[k][r] * A[k][c]);

  }
}
\end{Verbatim}
\end{minipage}
\quad 
\begin{minipage}[b]{0.48\textwidth}
\begin{Verbatim}
// With posit based accumulation
float AtA_f[2][2]; int r, c, k;
for (r = 0; r < 2; r++) {
  for (c = 0; c < 2; c++) {
    fn_init_p_quire (0);            // init quire
    // accumulation loop (using quire)
    for (k = 0; k < Arows; k++) 
       fn_posit_fma (A[k][r], A[k][c]);
    AtA_f[r][c] = fn_read_quire (); // read quire
  }
}
\end{Verbatim}
\label{code:lk}
\end{minipage}
  
The code sample in~\ref{code:lk} illustrates the modifications required to change a loop to use the quire instead of accumulating into a floating-point variable. The code on the left is the original loop that accumulates into float variables in memory and in the code on the right, the same loop has been rewritten to use the quire. Such  modifications could be the first step for an application developer to evaluate if posit based accumulation is suitable for their application. As explained in section~\ref{sec:clarinet-eg} this approach does introduce the overhead of converting between floating point and posit types.

In our simulation of optical flow on Clarinet we are using a set of three images, each 240x240 with a kernel size of 5x5 for the velocity computation. The cycle measurements focus on the velocity calculation routines which accounts for about 70\% of the total cycles run in the simulation.  There are three dot-product loops as part of the velocity computation. For the first two loops, the size of the vectors depends on the kernel size, and for a 5x5 kernel, the dot product operates on a vector of length 25. From simulations we observe that each dot product takes 548 cycles with posits versus 495 cycles with floats representing a 16\% overhead in cycles due to the use of posits, where each dot product involves 25 accumulations. However, the third dot-product operates on vectors of length 2, and while computing with float requires 25 cycles, this increases to 95 cycles with posits. Over the entire duration of the simulation, posits require 25.6 million cycles while float complete the same computation in 21 million cycles. This corroborates with our observations in~\ref{sec:blaslapack} where posits outperform floats in terms of latency when the number of accumulations in a loop is large enough to effectively pipeline the posit computations in the Melodica pipeline.

\subsection{Emulation}\label{sec:res}
\input{results}

%% file: results.tex
\begin{table}[!b]
\tiny
\centering
\vspace{-3mm}
\caption{Clarinet Experimental Configurations}
\begin{tabular}{|p{0.2cm} |p{2.1cm}| P{1.7cm} | p{1.1cm}| p{1.0cm}| p{1.5cm}| p{1.3cm}| p{1.2cm}|}
\hline
Sl. & Configuration           & Floating Point Unit& FP Registers & FP Divider \& Sqrt & Melodica       & Posit Registers & Posit Divider   \\
\hline
1   & Flute-F            & Single Precision   & 32 x 32-bit  & \xmark             & \xmark       & \xmark          & \xmark        \\
2   & Flute-F-DIV        & Single Precision   & 32 x 32-bit  & \cmark             & \xmark       & \xmark          & \xmark        \\
\hline
3   & Clarinet-F-P8.0       & Single Precision   & 32 x 32-bit  & \xmark             & N=8, es=0    & 32 x 8-bit      & \xmark        \\
4   & Clarinet-F-P16.1      & Single Precision   & 32 x 32-bit  & \xmark             & N=16, es=1   & 32 x 16-bit     & \xmark        \\
6   & Clarinet-F-P32.2      & Single Precision   & 32 x 32-bit  & \xmark             & N=32, es=2   & 32 x 32-bit     & \xmark        \\
\hline
7   & Clarinet-F-P8.0-DIV   & Single Precision   & 32 x 32-bit  & \cmark             & N=8, es=0    & 32 x 8-bit      & \cmark        \\
8   & Clarinet-F-P16.1-DIV  & Single Precision   & 32 x 32-bit  & \cmark             & N=16, es=1   & 32 x 16-bit     & \cmark        \\
10  & Clarinet-F-P32.2-DIV  & Single Precision   & 32 x 32-bit  & \cmark             & N=32, es=2   & 32 x 32-bit     & \cmark        \\
\hline
11  & Clarinet-P8.0           & \xmark             & \xmark       & \xmark             & N=8, es=0    & 32 x 8-bit      & \xmark        \\
12  & Clarinet-P16.1          & \xmark             & \xmark       & \xmark             & N=16, es=1   & 32 x 16-bit     & \xmark        \\
14  & Clarinet-P32.2          & \xmark             & \xmark       & \xmark             & N=32, es=2   & 32 x 32-bit     & \xmark        \\
\hline
15  & Clarinet-P8.0-DIV       & \xmark             & \xmark       & \xmark             & N=8, es=0    & 32 x 8-bit      & \cmark        \\
16  & Clarinet-P16.1-DIV      & \xmark             & \xmark       & \xmark             & N=16, es=1   & 32 x 16-bit     & \cmark        \\
18  & Clarinet-P32.2-DIV      & \xmark             & \xmark       & \xmark             & N=32, es=2   & 32 x 32-bit     & \cmark        \\
\hline
19  & Flute-D                 & Double Precision   & 32 x 64-bit  & \xmark             & \xmark       & \xmark          & \xmark        \\
20  & Flute-D-DIV             & Double Precision   & 32 x 64-bit  & \cmark             & \xmark       & \xmark          & \xmark        \\
\hline
\end{tabular}
\label{tab:clarinet_configs}
\end{table}

\begin{figure}[!t]
\centering
\subfloat[LUT utilization by sub-units in Melodica]{
    \includegraphics[width=0.45\columnwidth]{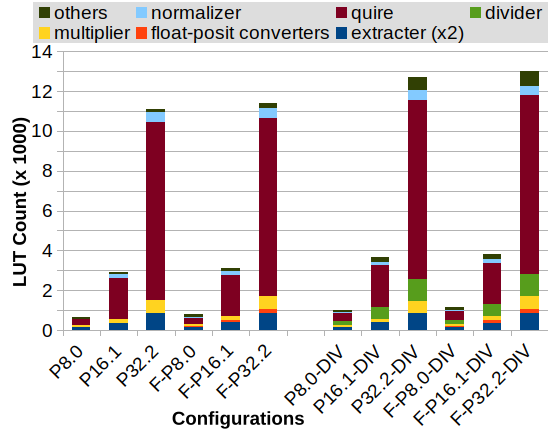} \label{fig:mel-synth-bar}}
\subfloat[LUT distribution across sub-units in Melodica]{
    \includegraphics[width=0.45\columnwidth]{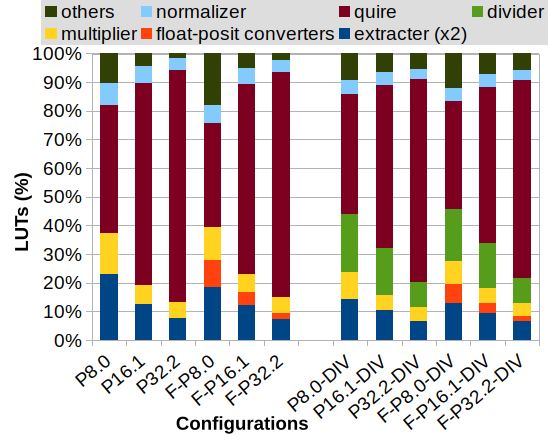} \label{fig:mel-synth-percent}}
\caption{Implementation Results for Melodica (100 MHz, Zed Board)}
\vspace{-5mm}
\end{figure}
\begin{figure}[!t]
\subfloat[LUT Utilization (absolute) by sub-units in Clarient CPU]{
    \includegraphics[width=0.40\columnwidth]{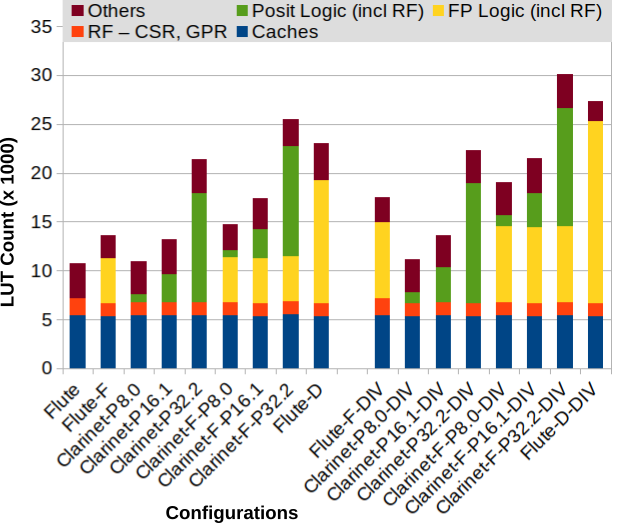} \label{fig:cpu-synth-bar}}
\subfloat[LUT distribution (\%) across sub-units in Clarinet CPU]{
    \includegraphics[width=0.55\columnwidth]{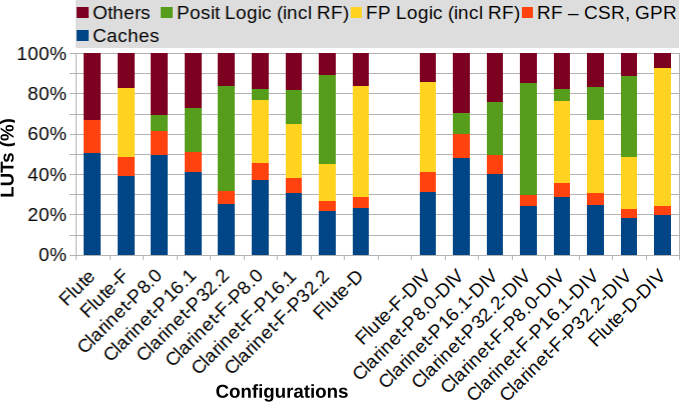} \label{fig:cpu-synth-percent}}
\caption{Implementation Results for Clarinet CPU (25 MHz, Zed Board)}
\vspace{-5mm}
\end{figure}

Clarinet was emulated on a Zed board with a \texttt{xc7z020-clg484-1} FPGA. We used Vivado 2020.2 for synthesis.
The CPU configurations chosen for emulation have been summarized in table~\ref{tab:clarinet_configs} and correspond to the configurations used to execute applications in sections~\ref{sec:blaslapack} and~\ref{sec:lk}.
Detailed implementation analysis was also done for the Melodica posit arithmetic unit as well.
For analyzing Melodica implementation, all the configurations in table~\ref{tab:clarinet_configs} with a posit arithmetic unit (configurations 3 to 18) were considered.

The bars in figure~\ref{fig:mel-synth-bar} represents the number of LUTs in the xc7z020-clg484-1 FPGA required to implement the different sub-unis comprising Melodicas across the configurations of interest.
The sub-units considered were (two) extractors, the normalizer, float-posit converters (in selected configurations), the multiplier, the divider (in selected configurations), and the quire.
The quire is easily the most resource intensive sub-unit across all configurations as seen clearly from figure~\ref{fig:mel-synth-percent} where the LUT utilization data is represented as a percentage of the total utilization for a particular Melodica configuration.
The size of the quire (in terms of LUTs utilized) grows non-linearly with posit-width (approaching $O(N^2)$, and this reflects in the sizes of the melodicas as well. At the CPU level all the different clarinet configurations listed in the Table~\ref{tab:clarinet_configs} were synthesized. While the complete configuration space of Clarinet is too large to exhaustively discuss here, a subset of configurations that were used to execute the applications listed in sections~\ref{sec:blaslapack} and~\ref{sec:lk} were selected as candidates for implementation analysis. The baseline for comparisons is configuration 1, a 32-bit RISC-V Clarinet processor with support for single-precision (32-bit) floating-point arithmetic (without FP divide and square root). No support for posits exist in this configuration.

%% file: conclusion.tex
We presented Clarinet -- an open-source, hardware-software framework for posit arithmetic empiricism. By integrating a parameterized posit arithmetic unit (Melodica) that supports quire-based fused operations with a RISC-V processor extended with special instructions for posit and quire operations, we created the Clarinet platform that enables experimentation around multiple axes related to posit arithmetic. Floating-point and posit data types can coexist in applications that run on Clarinet, allowing researchers to use it as a platform to gauge the impact of posits in real-world floating-point applications. We illustrated these capabilities through case studies on BLAS and LAPACK kernels and Lucas-Kanade optical flow estimation. We characterized Clarinet for latency, operating frequency, and resource utilization on Xilinx FPGAs across several configurations. In the future, we plan to explore Melodica's use as a posit-enabled accelerator.